\documentclass[11pt]{article}
\usepackage{graphicx}
\usepackage{amssymb,amsthm,amsmath,float}
\usepackage{epstopdf}

 \newcommand{\rem}[1]{}

\newcommand{\Eq}[1]{(\ref{#1})}

\newcommand{\Sec}[1]{\S \ref{#1}}
\newcommand{\Fig}[1]{Fig.~\ref{#1}}

\newcommand{\App}[1]{Appendix~\ref{#1}}

\newcommand{\mapformeqn}[9]{
			\begin{equation}\label{eq:#9}
			#1 : \left( \begin{array}{cc} #2 \\ #3 \end{array} \right) \mapsto
			\left( \begin{array}{cc} #4 \\ #5 \end{array} \right) =
			\left( \begin{array}{cc} #6 \\ #7 \end{array} \right) #8
			\end{equation}
			}

\newcommand{\InsertFig}[4]
{\begin{figure}[ht]
       \centerline{
         \includegraphics[width=#4]{./figures/#1}
       }
       \caption{{\footnotesize  #2}
       \label{#3}}
\end{figure}}

\newcommand{\InsertFigTwo}[5] {
\begin{figure}[ht]
       \centerline{
         \includegraphics[width=#5]{./figures/#1}
         \hskip 0.5in
         \includegraphics[width=#5]{./figures/#2}
       }
       \caption{{\footnotesize  #3}
       \label{#4}}
\end{figure}}
\newcommand{\InsertFigTwoHeight}[5] {
\begin{figure}[ht]
       \centerline{\includegraphics[height=#5]{./figures/#1} }
       \centerline{\includegraphics[height=#5]{./figures/#2} }
       \caption{{\footnotesize  #3}
       \label{#4}}
\end{figure}}

\newcommand{\R}{{\mathbb{ R}}}

\newcommand{\T}{{\mathbb{ T}}}
\newcommand{\Z}{{\mathbb{ Z}}}
\newcommand{\N}{{\mathbb{ N}}}

\newcommand{\cO}{{\cal O}}
\newcommand{\cL}{{\cal L}}
\newcommand{\cJ}{{\cal J}}
\newcommand{\eps}{\varepsilon}

\newcommand{\vphi}{\varphi}
\newcommand{\avJ}{\langle J_1 \rangle_\vartheta}

\newcommand{\rank}[1]{\mbox{rank}(#1)}

\begin{document}
\title{Normal Forms for Symplectic Maps with Twist Singularities}
\author{
     H.~R.~Dullin\thanks 
     {
	      H.R.Dullin@lboro.ac.uk. ,
	     Supported in part by DFG grant Du 302/2,and EPSRC grant GR/R44911/01.
     }
     ~and A~.V.~ Ivanov%
\\
     Department of Mathematical Sciences,\\
     Loughborough University \\
     Loughborough LE11 3TU, UK\\
\and
        J.~D.~Meiss\thanks
      {
	        James.Meiss@Colorado.EDU.
	        JDM was supported in part by NSF grant DMS-0202032.
	        The authors would like to thank the Mathematisches Forschungsinstitut Oberwolfach for
	        its support for a ``research in pairs" visit.
      }\\
     Department of Applied Mathematics\\
     University of Colorado \\
     Boulder, CO 80309-0526 \\
}
\maketitle
\begin{abstract}
\vspace*{1ex}
\noindent

We derive a normal form for a near-integrable, four-dimensional symplectic map with a fold or cusp singularity in its frequency mapping. The normal form is obtained for when the frequency is near a resonance and the mapping is approximately given by the time-$T$ mapping of a two-degree-of freedom Hamiltonian flow. Consequently there is an energy-like invariant.  The fold Hamiltonian is similar to the well-studied, one-degree-of freedom case but is essentially nonintegrable when the direction of the singular curve in action does not coincide with curves of the resonance module. We show that many familiar features, such as multiple island chains and reconnecting invariant manifolds, are retained even in this case. The cusp Hamiltonian has an essential coupling between its two degrees of freedom even when the singular set is aligned with the resonance module. Using averaging, we approximately reduced this case to one degree of freedom as well. The resulting Hamiltonian and its perturbation with small cusp-angle is analyzed in detail.

\end{abstract}

\section{Introduction}\label{sec:intro}
In a Liouville integrable Hamiltonian flow with $d$ degrees of freedom almost all motion takes place on invariant $d$-dimensional tori. The motion on these tori is conjugate to a linear flow with frequencies $\omega \in \R^d$. The linear flow is
$(\theta_0, t) \mapsto \theta = \omega t + \theta_0$
where $\theta \in \T^d$ is an angle. The map that assigns to each torus its frequency $\omega$ is called
the frequency map. For a Hamiltonian flow there are two natural frequency maps. In addition to the one just defined, the frequency-ratio map is obtained by fixing the energy and by turning the flow into a Poincar\'e first return map. The resulting map has $n = d-1$ degrees of freedom, and its frequencies are the $n$ frequency ratios $\omega_1 : \omega_2 : \dots : \omega_d$ of the original flow. In this way the frequency map of an integrable map can always be interpreted as the frequency-ratio map of some flow. Typically these are anharmonic---the oscillation frequency changes with the amplitude of oscillation. In Hamiltonian dynamics this property is called ``twist."  When this twist property breaks down, we say the frequency map has a singularity.  Twist singularities are are not unusual; for example, the frequency is not a monotone function of action in any system that has a pair of nested separatrices \cite{Morozov99, Morozov02}. This ``fold" singularity also generically occurs in any system near tripling resonances \cite{Moeckel90, Dullin99, Dullin03}. In this paper we will discuss both the fold  and cusp singularities and some of the dynamical consequences of the breakdown of twist. The cusp singularity corresponds to the collision of a pair of fold curves; it is another stable singularity when $n \ge 2$.

Consider the integrable symplectic map $ (\theta, J) \rightarrow (\theta + \Omega(J), J)$ with $\theta \in \T^n$ and $J \in \R^n$. Because the map is symplectic, the frequency map is {\em Lagrangian}, which means that $\Omega$ is the gradient of a scalar generating function $S(J)$. The twist corresponds to the Jacobian matrix $\tau = D\Omega$, or equivalently to the Hessian $D^2S$. There is a singularity in the frequency map when $\det(\tau) =0$ for some set of action values. Such a set is called the singular set and the corresponding tori are said to have {\em vanishing twist}. The image of the singular set is the {\em caustic}. A fold is one such singularity and it is stable for Lagrangian mappings; consequently, a one parameter family of such maps will not destroy the fold, but move it around in the frequency space. In particular, the caustic will cross rational frequencies. When the map is perturbed by a small periodic perturbation interesting dynamics is expected when the fold is near a resonant frequency such that the resonance is in the image of the frequency map. This situation is well studied for the case $n=1$ where the map is area-preserving and leads to a simple one-degree-of freedom Hamiltonian model \cite{Howard95, Simo98}. The canonical example is called the ``standard nontwist map,"  for reviews see \cite{Apte03, Wurm04, Wurm05}. In our study we will derive a generalization of this map for 
the four dimensional case.  

The transformation of the map to a normal form is inspired by singularity theory of Lagrangian mappings  \cite{Arnold85, Arnold88, Arnold95}. The class of transformations is, however, more restricted. The canonical variables of an integrable map are the angle-action  coordinates so the transformations should respect the periodicity of the angles. Moreover,  since we are in a near-integrable situation there is a natural distinction between momenta and coordinates which we would like to maintain as much as possible. These restrictions are quite severe, for example they restrict the choice of coordinate transformations to cotangent lifts of linear maps $A \in SL(n,\Z)$. In singularity theory, linear transformations are used to normalize the quadratic terms, while nonlinear transformations are used to remove most of the cubic terms. The latter is impossible in our setting, since we are restricted to linear transformations, but we can use scaling to bring out the singular behavior. Moreover, we cannot diagonalize the quadratic part using orthogonal transformations, since these are in general not in $SL(n,\Z)$. 

This will affect the normal form that we derive when the singular curve is not aligned with the resonance: we find it better to use noncanonical variables to decouple the required angle and action transformations. The resulting normal form is a Hamiltonian system, but one whose Poisson bracket is not in standard form.

After a brief review of the fold and cusp singularities in the context of integrable symplectic maps in \Sec{sec:singularities}, we will derive the normal form for a weakly nonintegrable mapping in \Sec{sec:normalForm}. The resulting normal form has a two degree-of-freedom Hamiltonian, shown in \Sec{sec:hamiltonian}. The dynamical behavior for the fold case and the cusp case are studied in the last two sections, \Sec{sec:fold} and \Sec{sec:cusp}.

\section{Fold and Cusp Singularities for Integrable Symplectic Maps}\label{sec:singularities}
An integrable, four-dimensional, symplectic map in angle-action coordinates has the form
\mapformeqn{F_{0}}{\vphi}{I}{\vphi'}{I' }{\vphi + \Omega(I)}{I}{,}{integrableMap}
where $(\vphi, I) \in \mathbb{T}^{2}\times \mathbb{R}^{2}$.

The dynamics of this system are completely encoded in the frequency map $\Omega: \mathbb{R}^{2} \to \mathbb{R}^{2}$, that defines the rotation vector on each invariant torus, $I = c$. This map is symplectic with the standard form $\varpi = d I_{1} \wedge d \vphi_{1} + d I_{2} \wedge d \vphi_{2}$ providing that the Jacobian matrix $D\Omega$ is symmetric.

The map \Eq{eq:integrableMap} is determined by a mixed-variable generating function 
\begin{equation} \label{eq:unperturbedGen}
   S_{0}(\vphi,I') = \langle \vphi, I' \rangle + T(I') \,,
\end{equation}
through the implicit equations $\vphi' = \partial S /\partial I'$ and $I = \partial S / \partial \vphi$.
Therefore, the gradient of the ``kinetic energy''  $T: \mathbb{R}^{2} \to \mathbb{R}$ gives the the frequency map, 
\begin{equation}
     \Omega(I) = DT(I)\,.
\end{equation}
Note that this implies that the Jacobian of $\Omega$ is symmetric since it is the Hessian of $T$.

In this paper we will study perturbations of \Eq{eq:integrableMap} for the case that the frequency map $\Omega$ is singular.

\subsection{Singularities of Lagrangian Maps}
In this section we briefly recall a few facts about the singularities of smooth
maps \cite{Arnold85}, ignoring the topology of the angle space.

A map $f:\R^{m}\rightarrow\R^{n}$ has a critical point at $x$ if its Jacobian, $Df(x)$, has less than maximal
rank, i.e., if $\rank{Df} < \min(m,n)$.  The image, $f(x)$, of a
critical point is a critical value.  A map is said to be ($C^{k}$)
stable at $x$ if every map that is sufficiently close to $f$ (in the
sense that the first $k$ derivatives are close) is locally
diffeomorphic to $f$.  The equivalence class of maps that are locally
diffeomorphic to $f$ is the ``germ'' of $f$.  If the dimension is low
enough, the germ can be represented by a fixed polynomial map, more
generally ``moduli,'' which are either parameters or arbitrary
functions, are needed to represent the germ.  The equivalence class of
maps represented by the germ at a critical point is called a
``singularity.''


A Lagrangian map is defined by the projection of a Lagrangian manifold
onto a Lagrangian plane.  A prime example is geometrical optics where
the Lagrangian manifold corresponds to a wave front together with its
unit normals---the velocity vectors, and the projection is to physical
space.  Correspondingly, the set of critical values of a Lagrangian
map is called a ``caustic.''  A Lagrangian manifold can be represented
by a single, generating function \cite{Weinstein77}; if, as in our
case, the Lagrangian manifold is a graph over the action space,
$(\vphi,J) = (DT(J),J)$, the generating function is $T(J)$.  The
Lagrangian map is the projection of the manifold onto the action
space, i.e. $\Omega: \R^{d} \rightarrow \R^{d}$ defined by $J \mapsto
DT(J)$.  The map has a critical point at $J$ if $D^{2}T(J)$ is
singular.

The standard theory of the singularities of Lagrangian maps has been
formalized by Thom and generalized by Arnold \cite{Arnold85}.  When
$d=1$ there is only one stable singularity, the ``fold.''  For $d = 2$
the ``cusp'' singularity is also stable.  For $d=3$, three new
singularities are stable, the ``swallowtail'' and two forms of point
singularities (pyramids and purses).

\subsection{Fold Singularity}
Suppose that $d=2$ and that the generating function has the formal power series expansion
\begin{equation}\label{eq:polyGen}
    T = \langle \omega^* , J\rangle + \sum_{i+j>1} t_{i,j} J_1^i J_2^{j}
\end{equation}
with the standard Euclidean scalar product $\langle, \rangle$.

The frequency map generated by $T$ is denoted
\[
    \Omega(J) = DT = (T_1,T_2)  = \omega^* + O(J) \;,
\]
where we denote the partial derivatives $\frac{\partial}{\partial J_i}$ with subscripts $i$. Thus $
\omega^*$ denotes the image of the origin $J=0$. The critical set is defined by the equation
\begin{equation}
	\tau(J_{1}, J_{2}) \equiv \det(D^2 T) = T_{11}T_{22} - T_{12}^{2} = 0 \;.
\end{equation} 
The set of critical points is a smooth codimension-one submanifold in $J$-space, i.e.~a curve, wherever the derivative $D\tau = (\tau_1, \tau_2) \neq 0$ on the critical set. In this case, coordinates can be chosen so that the critical set goes through origin, and is tangent to the $J_1$ axis, that is, so that $T_{12}(0) = T_{22}(0) = 0$. Under the nondegeneracy assumption that the Hessian $D^2 T$ has rank one,
\[
  T_{11}(0) = 2t_{2,0} \neq 0 \;.
\]
In this case, the coordinates can be scaled so that $t_{2,0}= \frac12$, and the generating function \Eq{eq:polyGen} becomes 
\begin{equation}\label{eq:polyGen2}
    T = \langle \omega^* , J\rangle + \frac12 J_1^2 +  \sum_{i+j>2} t_{i,j} J_1^i J_2^{j} \;.
\end{equation}

The singularity is locally a fold when the caustic is smooth---this occurs whenever the image of the tangent vector of the critical set is nonzero. A vector $v$ is tangent to the critical set when
\begin{equation}\label{eq:tangent}
  D\tau \cdot v = \tau_1 v_1 + \tau_2 v_2 = 0 \;,
\end{equation}
and its image, $D^2T \cdot v$, is nonzero if
$v$ is not in the kernel of $D^2T$,
\[
    \ker(D^2T) = \{ w : T_{11} w_1 + T_{12} w_2 = 0 \;,\; \tau = 0 \} \;.
\]
Therefore the function
\begin{align}\label{eq:foldcriterion}
    \kappa (J_1,J_2) & \equiv  T_{11}\tau_2 - T_{12}\tau_1 \nonumber\\
     & =  T_{11}(3T_{12}T_{122} - T_{11} T_{222}) + 
             T_{12}(T_{22}T_{111} -  3T_{12}T_{112})\;,
\end{align}
is zero only when the tangent is in the kernel, and the caustic is locally smooth when
\[
   \left. \kappa(J) \right|_{\tau(J) = 0}  \neq 0 \;.
\]
This is the nondegeneracy condition for the fold.

If we choose coordinates as in \Eq{eq:polyGen2}, so that the singularity crosses the origin parallel to the $J_1$ axis, then $\kappa(0) = -24 t_{0,3}$. Thus there is a fold at the origin if the coefficient of $J_2^3$ is nonzero. After suitable coordinate transformations, all other cubic terms can be eliminated, and the coefficients can be scaled to give the generating function
\begin{equation}\label{eq:foldgen}
    T_{fold} = \langle \omega^* , J\rangle + \frac12(J_1^2 + J_2^3) + O(4).
\end{equation}
Indeed, all higher order terms can be locally transformed away; hence, the generator \Eq{eq:foldgen} represents the germ of the fold \cite{Arnold85}. The fold singularity is denoted $A_{2}$.  

The frequency map of \Eq{eq:foldgen} is $\Omega(J) = \omega^* + (J_1, \frac32 J_2^2)$, and its critical set is the curve $J^*(s) = (s,0)$, determined by $\det(D^{2}T) = 3 J_{2} = 0$, which is the horizontal axis.  The caustic is the image of the critical set, 
$ \Omega^* = \Omega(J^*(s))$, which a horizontal line. 
The action of the frequency map is to fold the $J$-plane into the upper-half $\Omega$-plane.

\subsection{Cusp Singularity}\label{sec:integrableCusp}
If $\kappa = 0$ at some point on the critical set, then it is tangent to $\ker(D^2 T)$. The tangency is of first order if, for any tangent vector $v$, \Eq{eq:tangent},
\begin{equation}\label{eq:cuspcriterion}
   \mu(J) \equiv  D\kappa \cdot v  \neq 0\;.
\end{equation}
This is the nondegeneracy condition for the cusp singularity.

In the coordinate system of \Eq{eq:polyGen2}, a cusp occurs at the origin when $\kappa(0) = -24 t_{0,3}= 0$. In this case, the cusp nondegeneracy condition \Eq{eq:cuspcriterion} reduces to
\[
     \mu(0) = 96t_{1,2}(4t_{0,4}- t_{1,2}^2) \neq 0 \;.
\]
Consequently, the nondegeneracy conditions for a cusp are 
\[
   4t_{0,4} \neq t_{1,2}^2 \neq 0
\]
Coordinates for \Eq{eq:polyGen2} can then be chosen so that all of the cubic and quartic terms apart from $J_1J_2^2$ and $J_2^4$ can be transformed away; collecting terms gives rise to the form
\[
   T = \langle \omega^* , J\rangle+ \left(J_1 + \frac{1}{2} t_{1,2}J_2^2 \right)^2 + 
        \frac14 \left(4 t_{0,4}- t_{1,2}^2 \right)J_2^4 + O(5)
\]
After scaling, the cubic coefficient can be eliminated and the quartic one can be scaled to $\eps = \pm 1$, giving
\begin{equation}\label{eq:cuspgen}
    T_{cusp} = \langle \omega^* , J\rangle + \frac12 (J_{1}+J_{2}^{2})^{2} + \epsilon J_{2}^{4} +O(5)\;.
\end{equation}
As for the fold case, the higher order terms are irrelevant. There are two cusp singularities, depending upon the sign $\eps$ \cite{Arnold85}; they are denoted $A_{3}^{\epsilon}$. 
For \Eq{eq:cuspgen}, the critical set is the parabola $J^*(s) = (-(1+3\epsilon)s^2, s)$, 
and the caustic is the semi-cubical parabola, or cusp:
\[
    \Omega^* = \omega^*  -2\epsilon \begin{pmatrix}  3 s^2 \\  4 s^3
    \end{pmatrix} \;.
\]
In the exterior of the cusp, the map is one-to-one, while in the
interior it is three-to-one, see \Fig{fig:cuspplus} and
\Fig{fig:cuspminus}.  The preimage of the caustic consists of both
the critical set (shown in green in the figures) and the parabola $J_{1} =
-(1+\frac34 \epsilon)J_{2}^{2}$ (shown in red).  The tangent to the
critical set is vertical at the origin, and it is mapped to zero at
the cusp point.

When $\epsilon = 1$ the critical set is mapped to the caustic by
being, in effect, ``rotated'' about the negative $\Omega_{1}$ axis by
$180^{\circ}$, while the other preimage of the critical set is simply
squeezed, without rotation.

For $\epsilon = -1$, however, the critical set is simply ``squeezed''
towards the positive $\Omega_{1}$ axis, while the second preimage, a
parabola opening to the left, is rotated by $180^{\circ}$ and swept
around to the positive caustic curve.
      \InsertFig{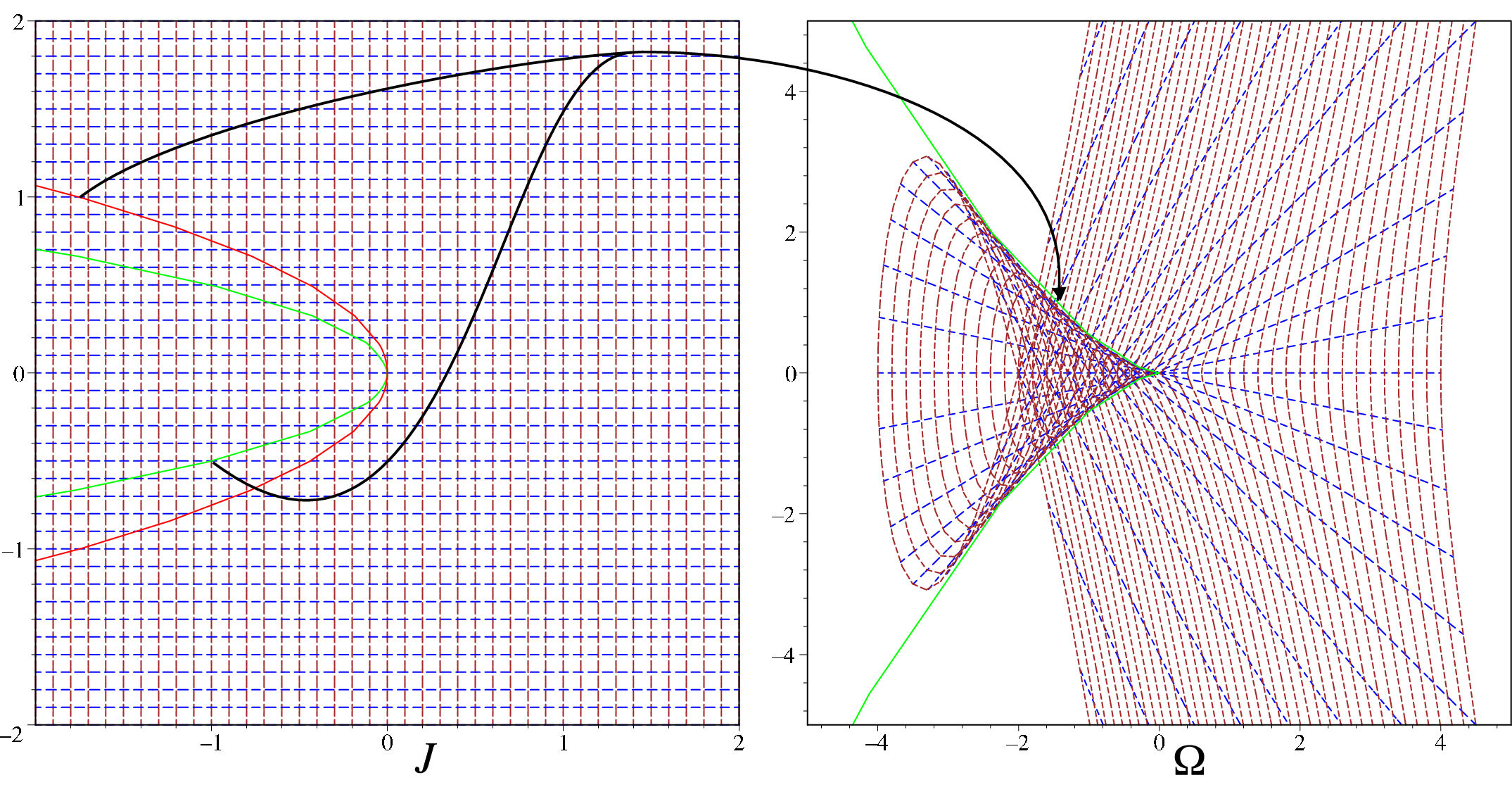} {The frequency map for the cusp
      $A_{3}^{+}$ given by \Eq{eq:cuspgen} with $\omega^*=0$. 
      The left figure shows the $J$-plane; the green
      parabola is the critical set, and the red is the second preimage of
      the caustic.  The right figure shows the $\Omega$-plane; the caustic is the
      green cusped curve.  The dashed lines show how a grid in $J$ is
      mapped to $\Omega$.} {fig:cuspplus}{5.0in}
      \InsertFig{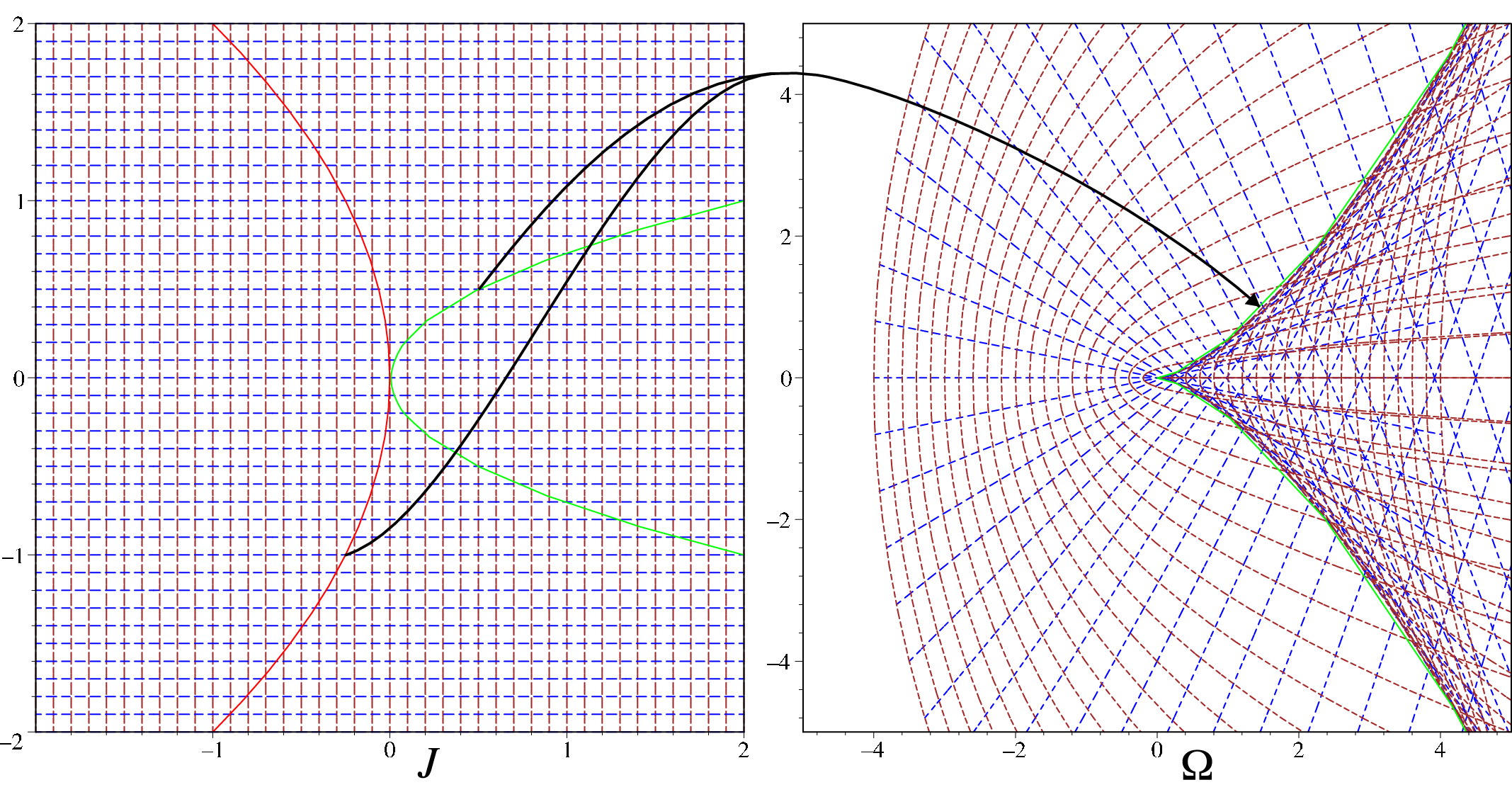} {The frequency map for the cusp
      $A_{3}^{-}$.  The designation of the curves is the same as
      \Fig{fig:cuspplus}.} {fig:cuspminus}{5.0in}


\section{Normal Form}\label{sec:normalForm}
In this section we will show how to unfold the fold and cusp singularities---that is, obtain families of symplectic maps,
\[
  (\vphi',I') =  F_{\eps,\delta}(\vphi,I)  \;,
\]
on $\T^2 \times \R^2$ that are perturbations of \Eq{eq:integrableMap} when $\Omega$ has a fold or cusp singularity. To do this requires (at least) two parameters that we will call $\delta$ and $\eps$. The first parameter will represent the unfolding of the singular set; it is the bifurcation parameter. We will assume that $\delta = 0$ corresponds to the case that the singularity is at a distinct location, namely when the fold or cusp passes through a ``resonance." The parameter $\eps$ represents the strength of the perturbation that destroys integrability. When the dependence on the parameters is not essential we shall suppress it in the notation. 

The family $F_{\eps,\delta}$ is generated by
\begin{equation}\label{eq:genFuncI}
   S(\vphi,I') =  \langle \vphi, I' \rangle + T_\delta(I') + 
              \eps S_1(\vphi, I', \eps, \delta)\;.
\end{equation}
Here we will assume that the perturbation $S_1$ is a smooth function of its arguments, and that 
$\Omega_\delta = DT_\delta$, has a fold or cusp singularity; thus, it has corank 1 on a critical set $I_\delta^*(s)$ with curve parameter $s$. The caustic is denoted
\[
	\Omega^*_\delta(s) = \Omega_\delta( I_\delta^*(s))
\]

The perturbed map will in general also have a perturbed frequency map, determined by the $\vphi$-independent portion of $S_1$. However, for simplicity we may assume that all of the changes in the frequency map are encoded by the parameter $\delta$; specifically, we assume
\[
     \int_{\T^2} d\vphi S_1(\vphi,I',\eps,\delta) = 0 \;.
\]     

Without loss of generality, the actions can be translated so that $I=0$ lies on the singular set, and in particular so that the origin corresponds to the point with the label $s=0$; thus we can assume that $I_\delta^*(0) = 0$, for any $\delta$, see \Fig{fig:foldSketch}. The origin maps to the point $\omega^*_\delta = \Omega^*_\delta(0)$ on the caustic.

%
\InsertFig{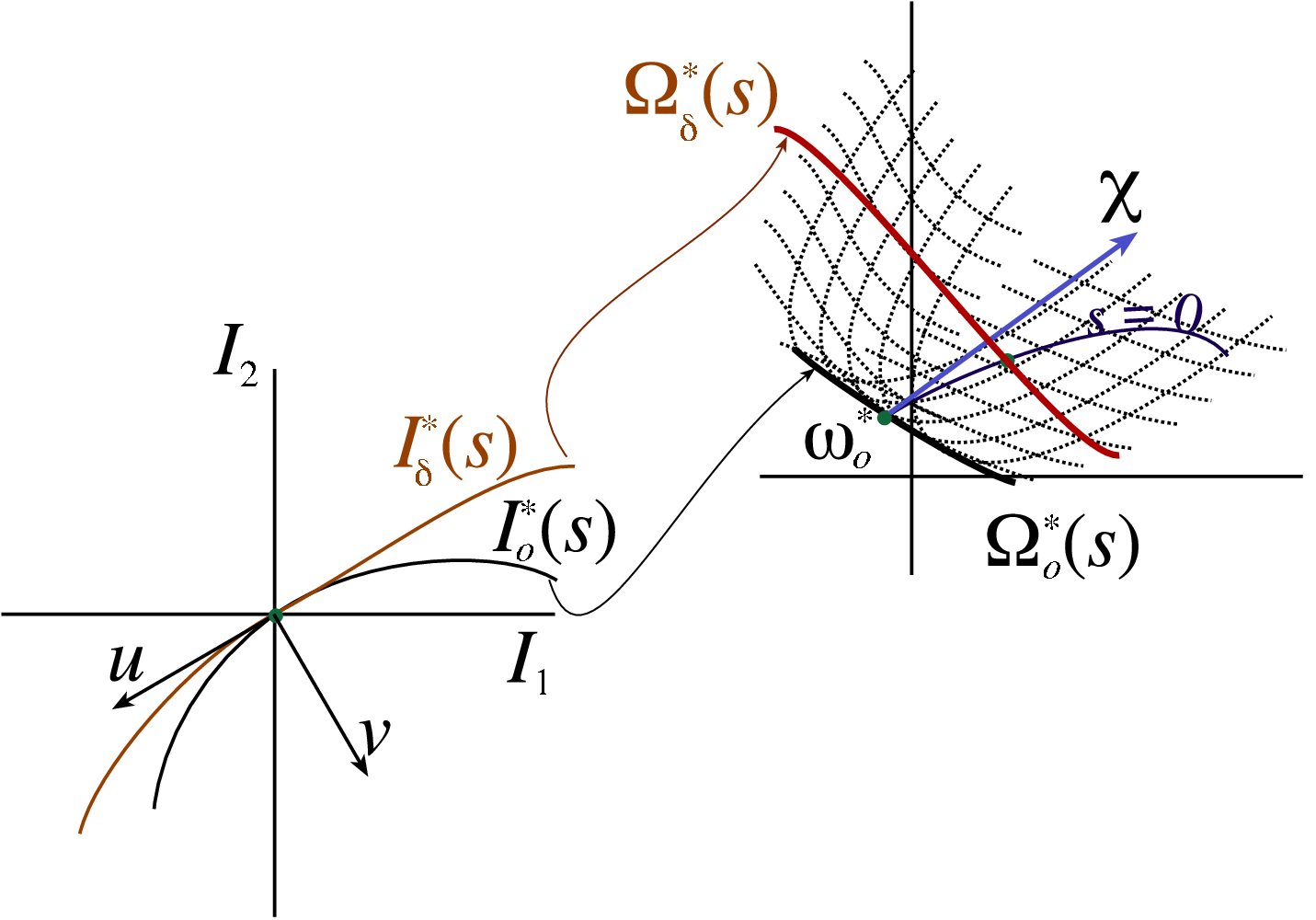}{Sketch of the $\delta$-dependent frequency map 
$\Omega_\delta$ for the case of a fold singularity. Here the origin in 
action space has been chosen so that $I=0$ is on the singular set and 
has constant slope.}{fig:foldSketch}{4in}
%

\subsection{Unfolding from Resonance}\label{sec:resonance}

A frequency $\omega^*$ is resonant if it is a rational vector:
\begin{equation}\label{eq:resonance}
    \omega^* = \left( \frac{n_1}{N}, \frac{n_2}{N}\right), \quad n_1, n_2 \in \Z, \quad N \in \N, 
       \quad {\rm gcd}(|n_1|,| n_2|, N) = 1 \,.
\end{equation}
More generally, for a given $\omega^*$, define the sublattice 
\[
    \cL = \left\{ m \in \Z^2 :   \langle m , \omega^* \rangle \in \Z \right\} \;.
\]
When $\dim{\cL} = 0$,  then we say $\omega^*$ is {\em incommensurate}, when $\dim{\cL} = 1$, $\omega^*$ is {\em commensurate}, and when $\dim{\cL} = 2$, $\omega^*$ is resonant. The sublattice for a resonant example is shown in \Fig{fig:sublattice}.
\InsertFig{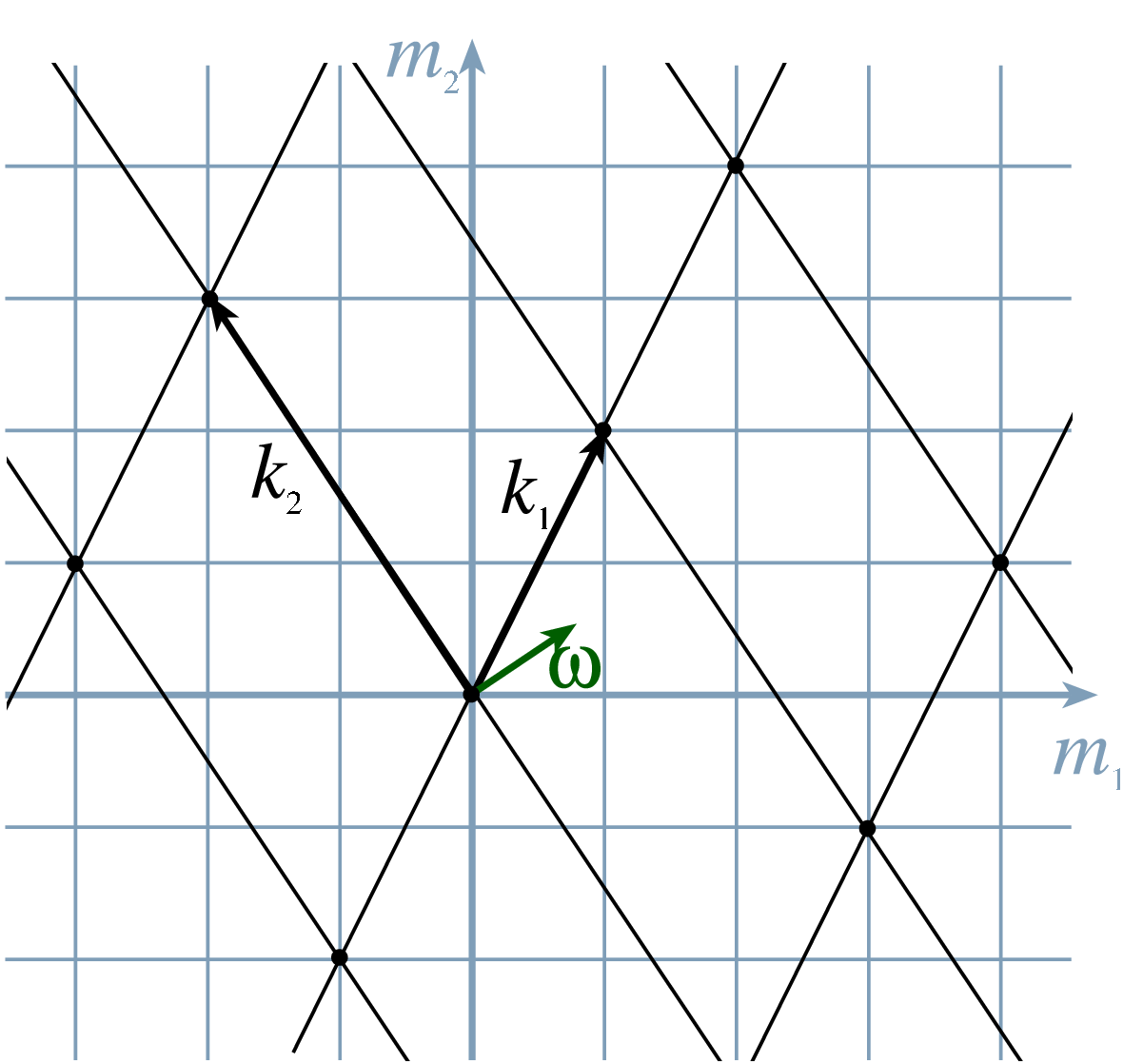}{Sublattice $\cL = A \Z^2$ for $\omega = (\frac37,\frac27)$ with $k_1=(1,2)$ and $k_2 = (-2,3)$. These basis vectors were chosen so that $k_1 \cdot \omega^* = {\rm gcd} (|n_1|, |n_2|) = 1$ and $k_2 \cdot \omega^* = 0$. $(k_1, k_2)$ is a minimal basis for the resonance module}{fig:sublattice}{3in}

We now argue that the parameters $(s,\delta)$ can be selected so that $\omega^* = \omega^*_0$ is resonant. This is particularly important to construct a near identity approximation to the mapping, see \Sec{sec:iteration}. First consider the fold case. The parameter $\delta$ unfolds the singularity if it causes the caustic $\Omega^*$ to ``move." For the fold this means that the map $\Omega^* :(s,\delta) \rightarrow \omega$ from a neighborhood of zero in $(s, \delta)$ to the neighborhood of $\omega^*$ in frequency plane is a local diffeomorphism.  This ensures that the fold actually crosses $\omega^*$ as $\delta$ varies, recall \Fig{fig:foldSketch}. Since rational points are dense in the frequency plane, we can then define the labels $s$ and $\delta$, so that $s=\delta = 0$ corresponds to a resonant
\begin{equation}\label{eq:omegaStar}
    \omega^* = DT_0(0) =  \Omega_0(0) \;.
\end{equation}
Note that we are still free to define the point $s=0$ on the singular sets for $\delta \neq 0$, and we will do so in the next section.

We next select coordinates to align the angle variables with the ``principle" resonance lines of $\omega^*$. Let $\{ k_1, k_2 \}$ be a basis of the resonant module $\cL$, recall \Fig{fig:sublattice}. Thus if we define the matrix 
\begin{equation}\label{eq:kdefine}
 A = (k_1, k_2) \;,
\end{equation}
then the sublattice is given by
\begin{equation} \label{eq:sublattice}
    \cL = \{ A k : k \in \Z^2 \} \;.
\end{equation}
There are a number of possible choices for these vectors; for example, they could correspond to the ``minimal" basis---that with shortest length. They could also be chosen to correspond to the largest resonant Fourier coefficients in the perturbation, i.e., the directions of the ``main" resonance lines that pass through $\omega^*$.  In the next section we will argue that they might be selected to approximately align with the slope of the singular set.

The matrix $A$ defines a transformation to new angle variables $\tilde \vphi = A^t\vphi$. A function $U$ on the original torus $\T^2$, i.e. a function that obeys the periodicity condition $U(\vphi + m) = U(\vphi), \forall m \in \Z^2$, is transformed into a function
\[
    \tilde U(\tilde \vphi) = U(A^{-1}\tilde \vphi)
\] 
on the new torus
\begin{equation}\label{eq:newTorus}
    \tilde \T^2 = \R^2 /
     A^{t}\Z^2 \;,
\end{equation}
that is, a function with the periodicity condition $\tilde U( \tilde \vphi + A^t m) = \tilde U(\tilde \vphi), \forall m \in \Z^2$, see \Fig{fig:newTorus}. In \Sec{sec:iteration}, we will use this periodicity to average the angle dependence and eliminate nonresonant Fourier terms in $\tilde{S}_1$. 
Generally the transformation $A$ is not unimodular 
\cite{Cassels71}; instead it has degree
$
     N = \det A \;,
$
meaning that $N$ copies of the new fundamental cell in $\tilde \vphi$ are covering the original torus in $\vphi$. Equivalently, the unit cell in $\tilde\vphi$ has area $\frac{1}{N}$.

\InsertFig{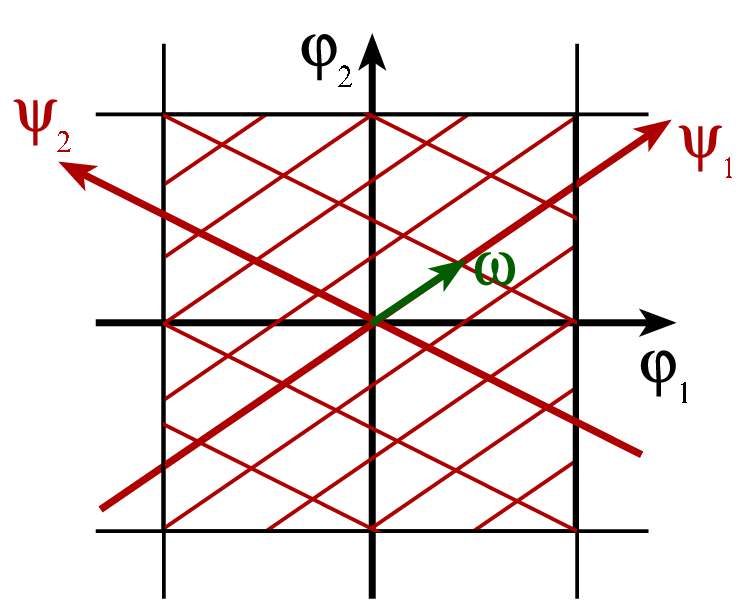}{Transformed angle coordinates in the resonance module for $\omega^* = (\frac37, \frac27)$ and $k_1 = (1,2)$, $k_2 = (-2,3)$ so that $\psi_1 = \vphi_1 + 2 \vphi_2$ and $\psi_2 = -2\vphi_1 + 3\vphi_2$. The black grid lines represent fundamental domains in $\vphi$, and the red, fundamental domains in $\psi$.}{fig:newTorus}{3 in} 

The angle transformation can be extended to a symplectic transformation, defining $(\tilde \vphi, \tilde I) = (A^t \vphi, A^{-1} I)$. As recalled in \App{sec:nonsymplectic}, the generating function for the map in the new coordinates under this ``point" symplectic transformation is simply $\tilde S(\tilde \vphi, \tilde I) = S(A^{-t}\tilde\vphi, A \tilde I)$, or
\begin{equation}\label{eq:genFuncII}
  \tilde S(\tilde \vphi,\tilde I') =  \langle \tilde \vphi, \tilde I' \rangle + 
      \tilde T_\delta(\tilde I') + \eps \tilde S_1(\tilde \vphi,\tilde I', \eps, \delta)
\end{equation}
where $\tilde T_\delta(\tilde I) = T_\delta(A \tilde I)$ and $\tilde S_1(\tilde \vphi, \tilde I, \eps, \delta) = S_1(A^{-t}\tilde \vphi, A \tilde I, \eps, \delta)$. Thus the new frequency is
\[
     \tilde \omega^* = D\tilde T_0(0) = A^t \omega^* \;.
\]

Suppose that $\tilde\chi$ is the first order change in the caustic with $\delta$, so that
\begin{equation}\label{eq:chidef}
    \tilde\omega_\delta^* = \tilde\Omega_\delta(0) = \tilde\omega^* +  \tilde\chi \delta + O(\delta^2) \;.
\end{equation}
The nondegeneracy condition ensures that $\tilde\chi \delta$ is not parallel to the fold---and freedom of choice of the point $s=0$ on the fold curves as $\delta$ varies implies that only the component of $\tilde\chi$ that is normal to the fold $\Omega^*(s)$ at $s=0$ is uniquely defined.  Below we will select the unfolding parameter to be this component of $\tilde\chi$. For the moment, we retain the parameter $\delta$ in the exposition for scaling purposes. 

For the cusp case,  it is natural to shift the actions so that the origin of action space maps to the cusp point, at least for the special values $s =0$, $\delta = 0$. Upon varying a single parameter $\delta$, the cusp moves along a curve in frequency space that generically does not go through any rational point, thus the cusp frequency is not generally resonant. If however, $\delta \in \R^2$ and the map $\Omega^*_\delta(0)$ is a local diffeomorphism from a neighborhood of $\delta = 0$ to a neighborhood of $\omega^*$,  then the origin $s = \delta =0$ can be selected so that the cusp frequency $\omega^*$ is resonant. 
For this case both components of $\tilde\chi \delta$ in \Eq{eq:chidef} are relevant. In the final normal form, we will use these components as the unfolding parameters.

\subsection{Diagonalization of the Kinetic Energy}\label{sec:diagonalization}
The second order terms in the kinetic energy correspond to a quadratic form $\tilde T_\delta^{(2)}(\tilde I) = \langle I, \tilde M_\delta I\rangle$, with ``mass matrix" $\tilde M_\delta$. By assumption $\tilde M_\delta$ has rank one, because the origin maps to the caustic. Let $v_\delta$ denote the unit vector along the kernel of $\tilde M$,
\[
  \tilde M_\delta v_\delta = 0 \;,
\]
and $u_\delta$ the unit vector orthogonal to $v_\delta $, i.e., the vector tangent to the singular set, recall \Fig{fig:foldSketch}. Since $\tilde M_\delta $ is symmetric,  this implies $\tilde M_\delta  = \lambda_\delta  u_\delta  u_\delta^t$ where $\lambda_\delta$ is its nonzero eigenvalue.  Thus the quadratic term in the kinetic energy can be written
\[
     \tilde T_\delta^{(2)}(\tilde I) = \lambda_\delta \langle u_\delta, \tilde I \rangle^2 \;.
\]

The vector $u_\delta$, tangent to the singular curve at the origin, will rotate with $\delta$. However, if the curvature of the singular set is nonzero at $s=0$, the label $s=0$ along the singular curve can be selected so that $u_\delta = u_0$. To see this, suppose that $u_\delta(s)$ is the tangent vector to the singular set at $s$. Then, according to the implicit function theorem, there is a local solution $s(\delta)$ to the equation $u_\delta(s) \times u_0 = 0$ precisely when $\frac{\partial }{\partial s} u_\delta(s)$ is nonzero at the origin in $(s,\delta)$. Relabeling the point $s=0$ on the singular sets for varying $\delta$ makes $u$ independent of $\delta$. We call this the {\em constant slope} normalization. Note that this normalization also applies to the cusp case, where $\delta \in \R^2$; however with this choice, the cusp is not necessarily the image of $I=0$ when $\delta \neq 0$.

Using the constant slope normalization, we can choose coordinates that are aligned with the kernel of $\tilde T_\delta^{(2)}$ once and for all. Letting $u^t = (\cos\alpha, \sin\alpha)$ be the unit vector tangent to the singular set, we define 
\begin{equation}\label{eq:actionRotation}
      \tilde I = R \hat{I} \;,  \quad     R = \begin{pmatrix}  \cos\alpha & -\sin\alpha \\ \sin\alpha & \cos\alpha \end{pmatrix}\;.
\end{equation}
With this transformation the new kinetic energy becomes 
\[
    \hat T_\delta (\hat{I}) \equiv \tilde T_\delta(R \hat{I}) = \langle \tilde\omega_\delta^* , R \hat{I} \rangle + 
                \lambda_\delta \hat{I}_1^2 + \tilde T_\delta ^{(3)}(R\hat{I}) + \ldots \;,
\]
so that the singular set is explicitly tangent to the $\hat{I}_1$ axis at the origin.
\rem{\footnote
{
   JDM: Using the constant slope normalization, we selected the origin so that 
   $u_\delta(0)$ is constant. For the fold case one might first choose (HRD: not always possible?)
   the point corresponding to $\delta = 0$ so that 
   the slope is rational. However, this cannot be done with a single parameter 
   $\delta$ while at the same time selecting a resonant $\omega^*$.
   In the cusp case, it make more sense to let $s = \delta = 0$ correspond 
   to the cusp point, and 
   there is no freedom left to select a slope value. 
}}

To make the transformation \Eq{eq:actionRotation} symplectic, it would be necessary also to transform the angle variables using $\hat{\vphi}  = R^t \tilde \vphi$ (recall \App{sec:nonsymplectic}); however, this is not a map on the torus. 
%
To get around this problem, we transform the map with the nonsymplectic transformation
$(\vphi, I) = (\hat{\vphi},R\hat{I})$, leaving the angle variables unchanged.
Under this linear transformation the generating function \Eq{eq:genFuncII} becomes
$\hat{S}(\hat{\vphi}, \hat{I'}) = \tilde{S}(\hat{\vphi}, R \hat{I'})$, or explicitly
\begin{equation}\label{eq:rotatedGen}
	  \hat {S}(\hat{\vphi}, \hat{I'})  =   \langle \hat{\vphi}, R \hat{I'} \rangle +
	   \hat T_\delta (\hat{I'}) + \eps \hat S_1(\hat{\vphi}, \hat{I'}, \eps, \delta)
\end{equation}

Since the coordinates are not symplectic, the generating function cannot be simply differentiated to get the map. Instead, the map is now defined using (recall \App{sec:nonsymplectic})
\[
    \hat{\vphi}' = R^{-t} \frac{\partial}{\partial \hat{I'}} \hat{S} \;, \quad
     \hat{I}      = R^{-1} \frac{\partial}{\partial   \hat{\vphi}} \hat{S} \;,
\]
which yields the implicit mapping
\begin{align}\label{eq:nonsymplecticMap}
    \hat{\vphi}' &=  \hat{\vphi} + \tilde\omega^* + 2\lambda \hat{I}_1 R^{-t} \hat{e}_1  + 
                           R^{-t}\nabla_{ \hat I '}(\hat{T}^{(3)} + \ldots + \eps \hat{S}_1) \;,\nonumber\\
     \hat{I}  & = \hat{I'} + \eps R^{-1} \nabla_{\hat{\vphi}} \hat{S}_1 \;.
\end{align}
Here the unit vector in the $\hat{I}_1$ direction is denoted by $\hat{e}_1$. 
This is a symplectic map written in non-symplectic coordinates.

If the slope $\tan\alpha$ were rational, then we could replace $R$ by a unimodular matrix whose second column is parallel to $v$ to accomplish the effect that we desire. In this case we could also transform the angle variables without destroying periodicity. Note that if the null vector $v$ is in the resonance module, then the same outcome could have been accomplished using the freedom in the selection of the matrix $A$, \Eq{eq:kdefine}, thereby avoiding the transformation \Eq{eq:actionRotation} altogether.\footnote
{
 If the null vector of the mass matrix in the original coordinate system $I$ is $v \in \cL$, we would choose $k_2 = v$, for then the transformed mass matrix $\tilde M = A^t M A$ has only one nonzero component.
} 

Even when the slope is not rational, we can select the module basis to nearly align with the null vector of $\tilde M$. In this case, the diagonalization can be accomplished with a rotation matrix $R$ with $\alpha$ small. Thus the general case can be considered to be that of small $\alpha$. Note however that the selection of  the basis $\{ k_1, k_2\}$ of the resonance module with very large integer coefficients will mean that the Fourier terms with small integer coefficients in the new basis may correspond very large coefficients in the original basis. For ``typical" functions, these coefficients will have small values and thus be less important dynamically.


\subsection{Scaling near Resonance}\label{sec:scaling}

To make explicit that we will study the neighborhood of the singularity at $\hat I=0$, we now choose a scaling so that $\hat I$ is small.  The goal is to scale the action so that the terms in the kinetic energy that give rise to the singularity are of order $\eps$, the same as the perturbation. Naively, this would be attained by assuming that $\hat I = \cO(\sqrt{\eps})$, i.e.,  using the transformation $(\hat \vphi, \hat I) = (\psi, \sqrt{\eps} J)$. Since this transformation is symplectic with multiplier (recall \App{sec:nonsymplectic}), the new generating function is obtained by the transformation ${S}(\psi, J) = \eps^{-\frac12}\hat S(\psi, \sqrt{\eps} J)$ or, from  \Eq{eq:rotatedGen},
\[
  {S}(\psi, J')  =   \langle \psi + \tilde\omega^*_\delta, R J' \rangle + 
                            \sqrt{\eps} \left[\lambda_\delta J'^{2}_1 + \hat S_1(\psi, 0,0, \delta)\right] + \cO(\eps) \;,
\]
showing that the quadratic term in the kinetic energy is now formally the same order as the perturbation. 

At this point the terms in $T^{(3)}$ that give rise to the fold normal form \Eq{eq:foldgen} are formally of higher order. Thus to ``see" the fold, we modify the transformation by differentially scaling the actions in the directions tangent and normal to the singular set. Using the scaling transformation 
\begin{equation}\label{eq:Bdefine}
      B = {\rm diag}( 1, \nu^{-1}) \;,
\end{equation}
we redefine
\begin{equation}\label{eq:scalingtrans}
   (\hat \vphi, \hat I) = (\psi,  \sqrt{\eps} B J )\;. 
\end{equation}
As before, the overall factor of $\sqrt{\eps}$ gives a transformation that is symplectic with multiplier, so that the new generating function is 
${S}(\psi, J) = \frac{1}{\sqrt{\eps}} \hat S(\psi, \sqrt{\eps} BJ)$, or
\begin{equation}
  {S}(\psi, J')  =   
         \langle \psi + \tilde\omega^*_\delta, R BJ' \rangle +    
         \sqrt{\eps}\left[ \lambda_\delta {J'_1}^2 +  \hat S_1(\psi, 0,0, \delta)\right]+
         \eps \nu^{-3} \kappa_\delta {J'_2}^3 + \eps \cO(1,\nu^{-1},\nu^{-2})
\end{equation}
When the coefficient $\kappa$ of $J_2^3$ is nonzero, the choice $\nu = \eps^{1/6}$ makes the cubic kinetic term $\cO(\sqrt{\eps})$ have the same order as the perturbation, and the neglected terms have higher order. Note that the nondegeneracy condition $\kappa \neq 0$ is precisely that for the fold \Eq{eq:foldcriterion}.

To enforce the near resonance condition not only the action, but also
the bifurcation parameter $\tilde\chi \delta$, must be small. We can set $\delta = \sqrt{\eps}$ to make this explicit, leaving $\tilde\chi$ to represent the unfolding parameter. It is also convenient at this point to introduce $\tilde\chi$ using \Eq{eq:chidef}, using the new representation
\[
      \chi \equiv (RB)^t \tilde\chi
\]
so that the components of $\chi$ correspond to the frequency shifts along and transverse to the fold, and the transverse component, $\chi_2$ is scaled with $J_2$. After this transformation we obtain
\begin{equation}\label{eq:scaledGenerator}
  {S}(\psi, J')  =   \langle \psi + \tilde\omega^*, R BJ' \rangle +    
                    \sqrt{\eps}\left[ \langle \chi , J' \rangle +{T}_{fold}(J) 
                    + U(\psi)\right] + \cO(\eps^{2/3})
\end{equation}
where $U(\psi) \equiv \hat S_1(\psi,0,0,0)$, and $ {T}_{fold}(J) = \lambda_0 {J}_1^2  +\kappa_0 {J}_2^3$.
Without loss of generality, we can scale the coefficients  so that 
\begin{equation}\label{eq:scaledFold}
    {T}_{fold}(J) = \frac12 ( {J}_1^2  +\kappa_0 {J}_2^3) \;
\end{equation}
by an overall choice of units for the action. The sign of the quadratic term can be 
changed by multiplying $S$ and $C$ by $-1$.

As was discussed in \Sec{sec:integrableCusp}, when $\kappa = 0$, the singularity is generically a cusp. In this case, the same scaling transformation \Eq{eq:Bdefine} to select the leading order terms in the kinetic energy can be used, though with the choice $\nu = \eps^{1/4}$. There are now two 
additional terms that are of order $\sqrt{\eps}$, one cubic and one quartic. 
The generating function again has the form \Eq{eq:scaledGenerator}, though the 
kinetic energy becomes
\rem{\footnote
	{ JDM: Possibly give the formulae for $\kappa$, $\rho$ and $\mu$? }
}
\begin{equation}\label{eq:scaledCusp}
   {T}_{cusp}(J) =  \frac12( {J}_1  + \rho_0 {J}_2^2)^2  + \mu_0 {J}_2^4 \;,
\end{equation}
and  the first corrections are now $\cO(\eps^{3/4})$.

Since the additional transformation \Eq{eq:scalingtrans} is 
also nonsymplectic, the map generated by $S$ is obtained from the relations
\begin{equation}\label{eq:nonsymplecticGenerator}
    \psi' = C^t \frac{\partial}{\partial J'} {S} \;, \quad
     J     = C   \frac{\partial}{\partial   \psi} {S} \;,
\end{equation}
where
\begin{equation}\label{eq:cDefine}
   C \equiv (RB)^{-1}  = \begin{pmatrix} \cos\alpha & \sin\alpha \\ -\nu\sin\alpha & \nu \cos\alpha  \end{pmatrix}\;.
\end{equation}
One clear advantage of the scaling is that the resulting map is now easily 
written explicitly or in the more compact leap-frog form
\mapformeqn{F_{\eps,\chi}}{J}{\psi}{J' }{\psi'}
		{J -\sqrt{\eps} C\nabla U(\psi)}
        {\psi + \tilde\omega^* + \sqrt{\eps}C^t \left( \chi + \nabla  T(J')\right) }
        {.}{perturbedMap}

The complete set of transformations from the original variables $(\vphi, I)$ of \Eq{eq:genFuncI} to the variables $(\psi,J)$ of \Eq{eq:perturbedMap} is
\begin{align}
     \psi &= A^{t}\vphi \;,\nonumber \\
      J   &= \frac{1}{\sqrt{\eps}} C A^{-1} I \;.
\end{align}

\subsection{Iterating near Resonance}\label{sec:iteration}

The map $F_{\eps,\chi}$ \Eq{eq:perturbedMap} becomes a near-identity map after it is iterated $N$ 
times, where $N$ corresponds to the denominator of the resonant frequency 
$\omega^*$, recall \Eq{eq:resonance}. The point is that, all of the change in \Eq{eq:perturbedMap} is $\cO(\sqrt{\eps})$, except for the term $\tilde \omega^*$. However, after $N$ iterations, this term becomes $N \tilde\omega^* = A^{t} N\omega^* \in A^{t}\Z^2$; therefore, this term is ``integral" on the transformed torus and consequently disappears from the phase map after iteration.Since $\sqrt{\eps}$ is assumed to be small, a near-identity map is obtained. 

More precisely, using \Eq{eq:perturbedMap}, the evolution is
\begin{align*}
    J_j &= J_0 - \sqrt{\eps} C \sum_{t=0}^{j-1} \nabla U( \psi_t ) \,, \\
 \psi_j &= \psi_0 + j  \tilde\omega^* + 
        \sqrt{\eps}C^t \left( j \chi + \sum_{t=1}^j \nabla T(J_t) \right) \;.
\end{align*}
Therefore $\psi_j =\psi_0+ j\omega^* + \cO(\sqrt{\eps})$, and the action after $N$ steps can be simplified to
\[
       J_N = J_0 - \sqrt{\eps} C \sum_{t=0}^{N-1} \nabla U( \psi_0 + t\tilde\omega^* ) + O(\eps) \,.
\]
This can be simplified by expanding the potential in a Fourier series. By assumption the perturbation is smooth, and thus so is the perturbing potential $U(\psi) = S_1(A^{-t} \psi, 0,0,0)$. This function is periodic on the transformed torus $\tilde \T^2$, \Eq{eq:newTorus}; therefore it has the expansion
\begin{equation}
        U(\psi) = \sum_{m \in \Z^2} c_m e^{2\pi i \langle A^{-1}m , \psi\rangle} \;.
\end{equation}
Using the identity
\[
   \sum_{l=0}^N e^{2\pi i l p/N} = \left \{ \begin{aligned} 
                   & 0 \text{ if } p/N \not \in \Z \\
                   & N \text{ else }
   \end{aligned} \right. \;,
\]
it is easy to see that only the resonant Fourier terms, those with $m \in \cL$, \Eq{eq:sublattice},
contribute to the Fourier series of $J_N$, so that
\[
  J_N = J_0 - \sqrt{\eps} N  C\nabla \bar U( \psi_0 ) + O(\eps) \,.
\]
Here $\bar U$ denotes the Fourier series of $U$ with all the nonresonant terms removed:
\begin{equation}
    \bar U(\psi) \equiv \sum_{ n \in \Z^2} c_{(An)} e^{i \langle  n, \psi \rangle } \;.
\end{equation}
Note that since all of the lower frequency terms have been removed from the Fourier series, $\bar U$ is now a function on the ordinary torus:
\[
  \bar U(\psi + k) = \bar U(\psi) \;, \quad \forall k \in \Z^2
\]
Thus, as far as the $N^{th}$ iterate is concerned, we can take $\psi \mod 1$.
Consequently, the term $N \tilde \omega^*$ in the $N^{th}$ iterate of the angle is equivalent to $0$, and since $J_j$ changes only at $\cO(\sqrt{\eps})$, we obtain
\[
   \psi_N = \psi_0 + \sqrt{\eps} N C^t \left( \chi + \nabla T(J_N) \right) 
   + O(\eps) \,.
\]
We can conclude by noting that the near identity map for the $N^{th}$ iterate has the generating function 
\begin{equation}\label{eq:finalGenerator}
    S_{\eps,\chi}(\psi,J') = \langle \psi, C^{-1}J'\rangle + \sqrt{\eps} N \left( 
    \langle \chi, J' \rangle + T(J') + \bar U(\psi) \right)  + O(\eps) \;.
\end{equation}
The map is generated from \Eq{eq:finalGenerator} using the nonsymplectic form \Eq{eq:cDefine}.
\section{Hamiltonian Normal Form}\label{sec:hamiltonian}

The main advantage of the near-resonance condition is that the $N$-step map generated by \Eq{eq:finalGenerator} is a near identity map; consequently, it is---to order $\eps$---the time $N \sqrt{\eps}$ map of a flow. This flow is generated by the Hamiltonian 
\begin{equation}\label{eq:Hamiltonian}
   H(\psi, J) = \langle \chi, J \rangle + T(J) + \bar U(\psi) \,.
\end{equation}
Here $T$ is either the fold, $T_{fold}$ \Eq{eq:scaledFold}, or the cusp, $T_{cusp}$ \Eq{eq:scaledCusp}, normal form. For the Hamiltonian, the nonsymplectic-form \Eq{eq:nonsymplecticGenerator} becomes a noncanonical Poisson bracket in $z = (\psi, J)$, and the equations of motion are given by
\begin{equation}\label{eq:nonsymplecticForm}
   \dot{z} = \cJ \nabla H \;, \quad 
   \cJ = \begin{pmatrix} 0 & C^{t} \\ -C & 0  \end{pmatrix}
\end{equation}
where $\cJ$ is the Poisson matrix.
The Hamiltonian flow preserves the energy, consequently the original map has an approximate conserved quantity when $\eps$ is sufficiently small. The existence of a conserved quantity for a 4D symplectic map is nontrivial.

The equations of motion for \Eq{eq:Hamiltonian}  are 
\begin{align*}
	\dot \psi & =   C^t( \chi +  \nabla  T(J) ) \\
	\dot J    & =  -C\nabla \bar U(\psi) 
\end{align*} 
Using the form \Eq{eq:cDefine} for $C$, these equations can be written
\begin{align} \label{eq:flow}
	\dot \psi_1     & =    \cos \alpha (\chi_1+  \partial_{J_1} T) - \nu  \sin\alpha ( \chi_2 +  \partial_{J_2} T) \nonumber\\
	\dot \psi_2     & =    \sin \alpha (\chi_1+  \partial_{J_1} T) + \nu  \cos\alpha ( \chi_2 +  \partial_{J_2} T) \nonumber \\
	\dot J_1    & =  -\cos\alpha \partial_{\psi_1}  \bar U - \sin\alpha \partial_{\psi_2}  \bar U  \nonumber \\
	\dot J_2    & =  \nu \left( \sin\alpha \partial_{\psi_1} \bar U - \cos\alpha \partial_{\psi_2} \bar U \right)
\end{align}
It is tempting to introduce new angles $\tilde \psi = R^t \psi$, 
but since the slope $\tan \alpha$ is generically not rational this transformation does not respect the periodicity of $\bar U$, with the result that $\bar U$ would be quasiperiodic function of the new angles.


The model \Eq{eq:flow} has several important parameters: the frequency mismatch vector $\chi$, the direction of the fold $\alpha$, and the perturbation size $\nu$. In addition, the kinetic energy for the fold model depends upon the relative size of the fold term, $\kappa_0$ and  cusp model has the coefficients $\rho_0$ and $\mu_0$. 
In addition there is of course the freedom of choice of the periodic potential $\bar U$. 
Note that $\nu$ is either $\eps^{1/6}$, or $\eps^{1/4}$ and is assumed small. If we take this to the extreme $\nu = 0$ then \Eq{eq:flow} implies that $J_2$ is a second constant of motion. Thus in some sense our system reduces to one degree of freedom. However, the angle dependence is quasiperiodic. We will discuss this case in \Sec{sec:quasiperiodic}.

\section{Fold Model}\label{sec:fold}
The system \Eq{eq:flow} can be further simplified for the fold model where $T = T_{fold}$ is given by  \Eq{eq:scaledFold}. In this case $\chi_1$ is irrelevant and can be set to zero without loss of generality by, for example, a shift along the $J_1$ axis. The point is that $\chi_1$ corresponds to a frequency shift along the fold, but in the normal form the fold is a straight line. The parameter $\chi_2$ is the essential detuning parameter.

If we assume that the original potential has strong Fourier components in the direction of the fundamental resonances, then the dominant terms in the potential will be $\bar U = -a \cos(\psi_1) - b \cos(\psi_2)$.\footnote
{
	It is also possible that the original potential has no terms in the resonance module so that 
	$\bar{U} = 0$. In this case a higher order normal form must be used. This is the situation 
	for some resonances (those with even period) in the standard nontwist mapping 
	\cite{Petrisor01, Wurm04, Wurm05}.
}
 In this case the Hamiltonian  \Eq{eq:Hamiltonian} becomes explicitly
\begin{equation}\label{eq:foldHam}
      H_{fold}(\psi, J) =  \chi_2 J_2 +  \frac12 (J_1^2 + \kappa_0 J_2^3) - a \cos(\psi_1) - b \cos(\psi_2) \,.
\end{equation}
Though this Hamiltonian appears to be separable into two individual one-degree-of-freedom systems, this is not the case when $\alpha \neq 0$ because of the nonsymplectic form \Eq{eq:nonsymplecticForm}.

It may be the case that additional terms in the resonance module will have amplitudes comparable to the ones included in \Eq{eq:foldHam}. In this case, these terms should be included in $\bar U$ and the dynamics of the two degrees of freedom are coupled even when $\alpha = 0$. 

The dynamics of the two-degree-of-freedom system \Eq{eq:foldHam} can be visualized using an appropriate Poincar\'e section. Though the standard sections are not complete, they can still give insights into the dynamics. For our purposes, it seems that the most useful section is the surface $\Sigma = \{ \psi_1 = 0, H = E\}$. This section can be projected onto the coordinates $(\psi_2, J_2)$ with a choice of sign for $J_1$, since the Hamiltonian \Eq{eq:foldHam} depends quadratically on $J_1$. Two 
example phase portraits are shown in \Fig{fig:foldsection1}. 

\InsertFigTwo{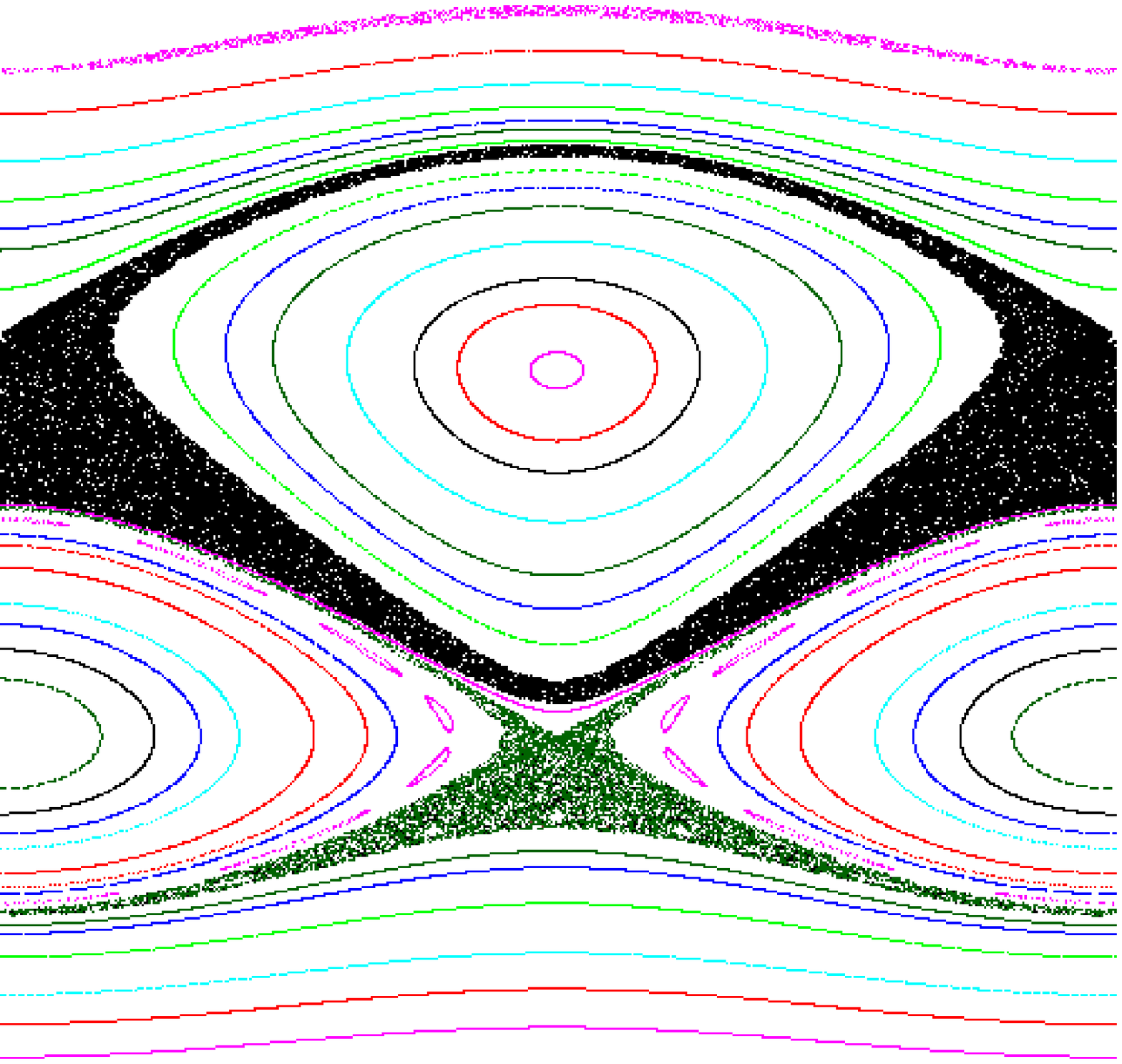}{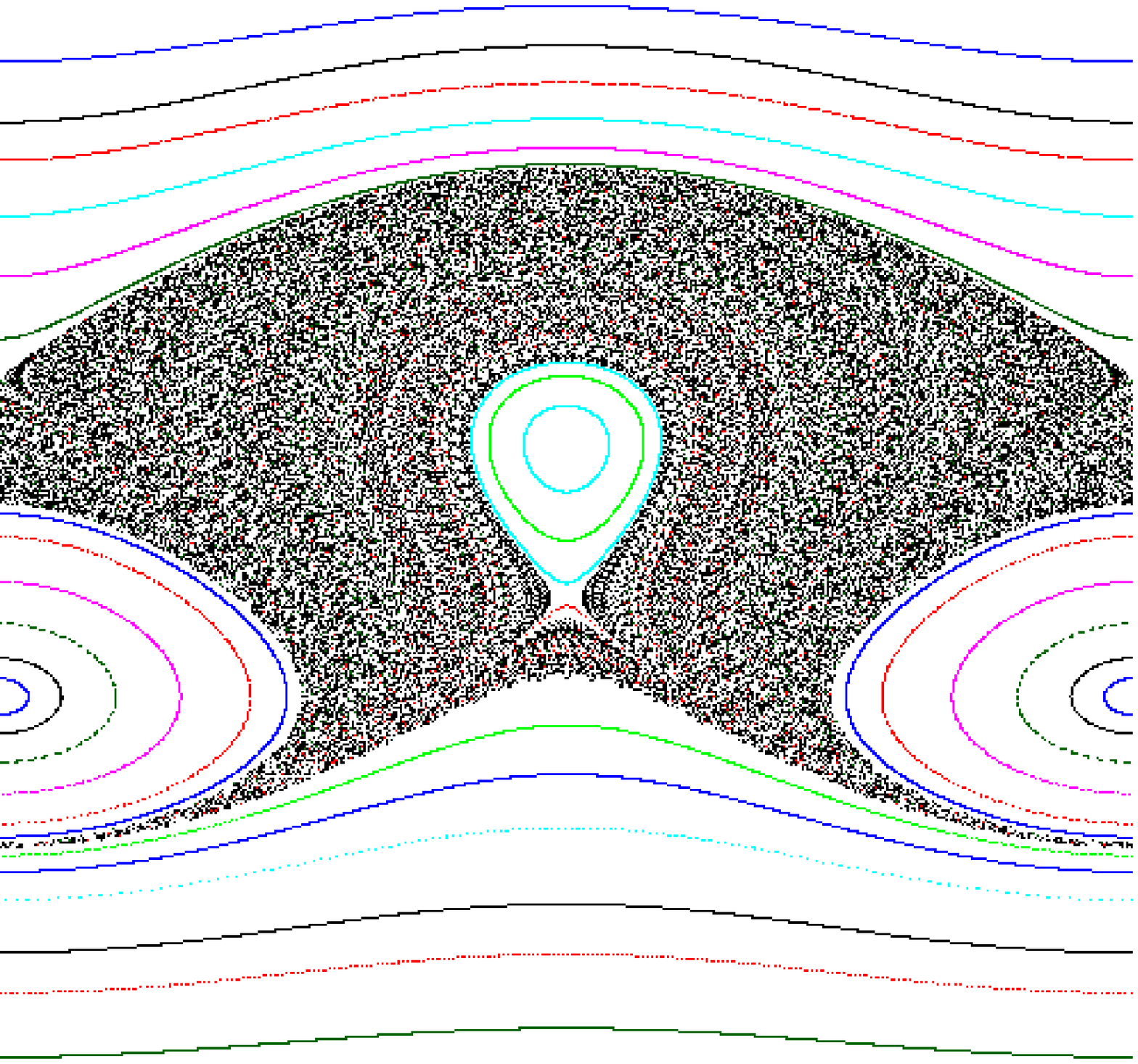}{Poincar\'e section  for the Hamiltonian \Eq{eq:foldHam} with $(\psi_2,J_2) \in \left[-\pi,\pi \right] \times \left[-3,2.5 \right]$ and $J_1 > 0$. The parameters are set to $a = b = \kappa_0 = 1$,  $E = 2$,  $\alpha = 0.01\pi$, and $\nu = 0.05$.  For the left panel $\chi_2 = -2.5$, and for the right $\chi_2 = -1.7$}{fig:foldsection1}{0.4\textwidth} 

In these sections there are two visible elliptic fixed points; they correspond to periodic orbits of the flow. For example in the right panel the orbit at $(\psi_2,J_2) = (0, -0.016)$ has motion in which $J_1$ is oscillating in narrow region about $2.5$ while $\psi_1$ rapidly rotates.  The orbit at $(\psi_2,J_2) = (\pi, -1)$ has its first degree of freedom on an oscillatory trajectory about the origin, with $J_1$ oscillating between $\pm 1.4$. For both of these orbits $\psi_2$ and $J_2$ remain nearly fixed. 

Our goal is to explain the main features of these phase portraits.

\subsection{Aligned Fold: $\alpha = 0$}
The model \Eq{eq:foldHam} reduces to the one-dimensional case if the fold happens to be aligned with the resonant module, so that $\alpha = 0$, for then the equations \Eq{eq:flow} decouple. The dynamics in $(\psi_1,J_1)$ then correspond to a simple pendulum.  In the decoupled $(\psi_2,J_2)$ dynamics, the parameter $\nu$ (if it is nonzero) can now be eliminated by rescaling time. The resulting canonical one-degree-of-freedom model,
\begin{equation}\label{eq:1doffoldmodel}
     H_{2}(\psi_2,J_2) = \chi_2 J_2 + \frac{\kappa_0}{2} J_2^3 - b \cos (\psi_2) \;,
\end{equation}
is the same as that originally studied by Howard \cite{Howard95} and derived as a normal form by Simo \cite{Simo98}. This system has equilibria at the points
\[
      \psi_2 = 0,  \pi  \;, \;  J_2^{(\pm)} = \pm \sqrt{ - \frac{2\chi_2}{3\kappa_0}} \;.
\]
For definiteness, suppose that $\kappa_0, b > 0$. When $\chi_2  < 0$ there are four equilibria as shown in \Fig{fig:foldphasespace}--two  are centers ($(0, J_2^{(+)})$ and $(\pi,J_2^{(-)}))$, and two are saddles ($(0, J_2^{(-)})$ and $(\pi,J_2^{(+)})$). These orbits correspond to the two families of resonant islands with rotation number $\omega^*$ in the original coordinate system that are caused by the fold. These resonant islands are destroyed in saddle-center bifurcations at $\chi_2 =0$. The separatrices of the saddles undergo a reconnection as  $\chi_2$ passes through
\begin{equation}\label{eq:chistar}
 \chi^* = -\frac32 (b^2\kappa_0)^\frac13 \;.
\end{equation}
When $ \chi^* < \chi_2 < 0$, there are ``meandering" invariant circles, i.e., invariant circles that go above the upper island chain and below the lower one.

\InsertFigTwoHeight{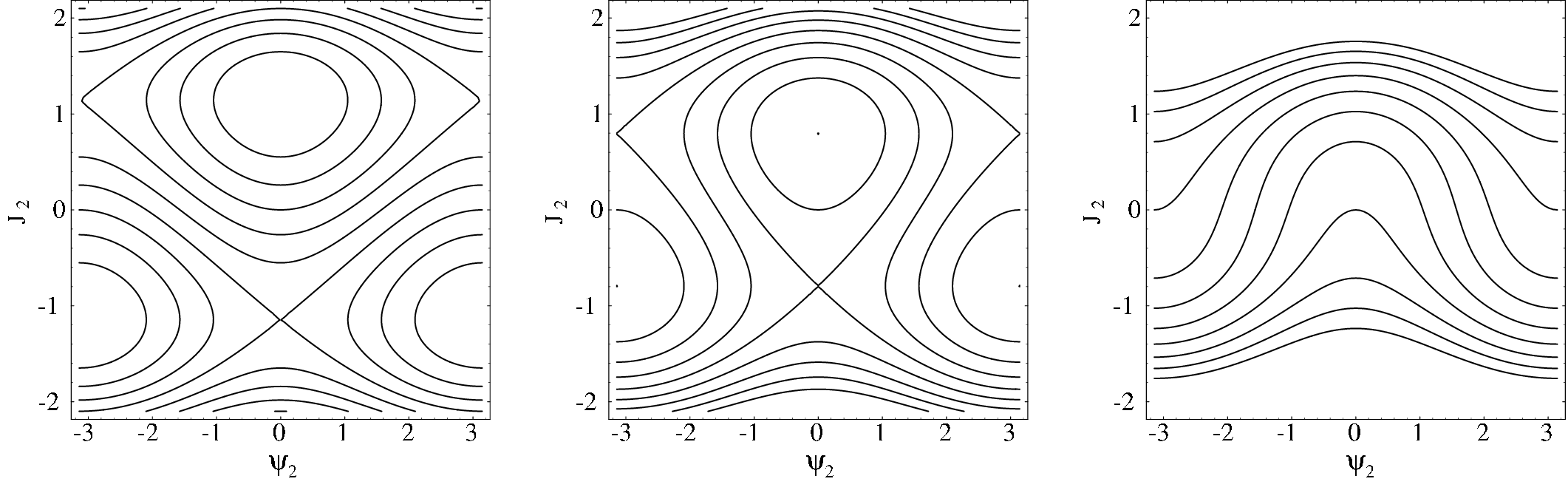}{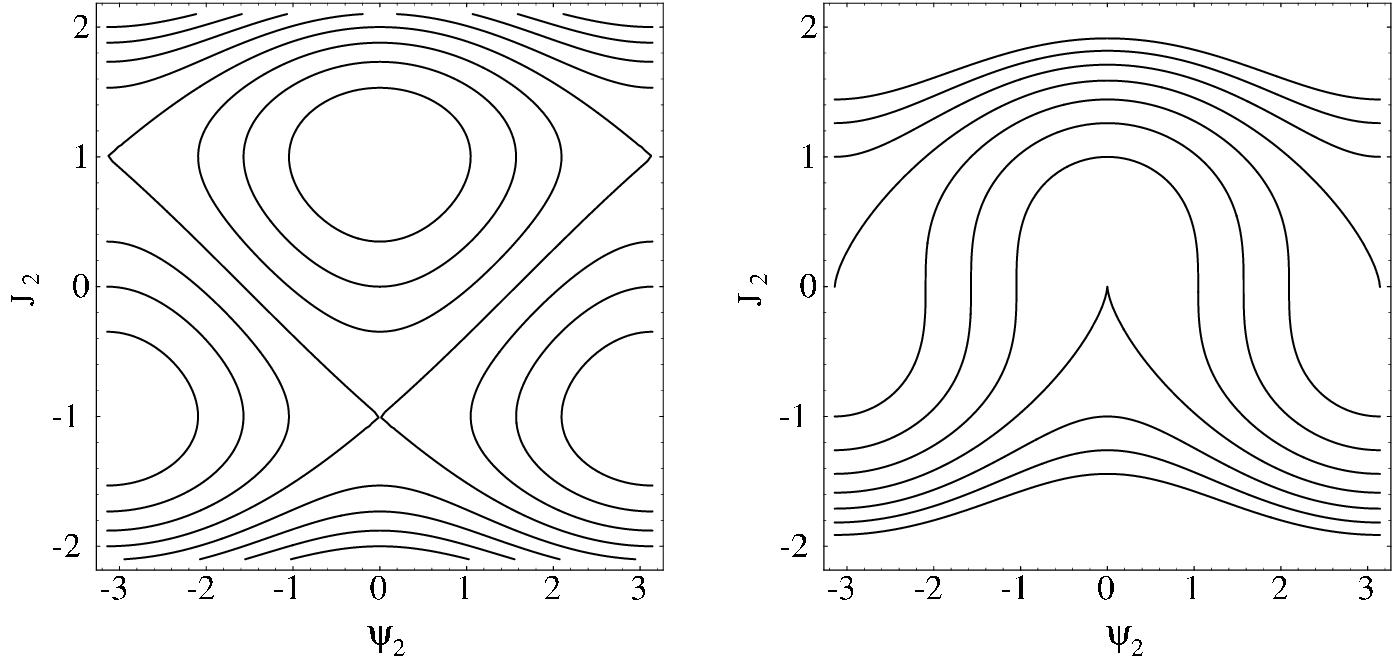}{Phase space of the one-degree-of-freedom fold model, 
\Eq{eq:1doffoldmodel} for $\kappa_0 = b = 1$ and $\chi_2 =$. The top line shows generic 
phase portraits for $\chi_2 = \chi^*(1/2)^{2/3}, \chi^*(3/2)^{2/3}, -\chi^*(1/6)^{2/3}$.
The center phase portrait shows a meandering curve.
The bottom line shows critical phase portraits for $\chi_2 = \chi^*, 0$. Contourlines shown are $H_2 = i/2$ 
for $i = -5, -4, \dots, 5$.}{fig:foldphasespace}{0.25\textwidth} 

The Poincar\'e sections for the full Hamiltonian \Eq{eq:foldHam} in \Fig{fig:foldsection1} are similar to the phase portraits in \Fig{fig:foldphasespace}, though since $\alpha \neq 0$ in the former figure the two degrees of freedom are coupled and some orbits are chaotic.  Since $\nu \ll 1$ the dynamics of  $J_2$ is much slower than that of $J_1$, and since $\alpha$ is also small, the coupling between the two is weak. It is interesting to note that the chaotic orbits in \Fig{fig:foldsection1} (in the regime of the meandering tori) correspond to values of $(\psi_1, J_1)$ that are near the pendulum separatrix for this degree of freedom. Conversely near the center of the islands on the section the corresponding values of $J_1$ are either large, so that the pendulum is in a rotating regime, or near zero so that the pendulum is deeply trapped. The island near $\psi_2=0$ is created in saddle-center bifurcation that occurs $\chi_2 \approx -1.47$ for the value $\alpha = 0.01\pi$ shown. By contrast, the saddle-center bifurcation for the island near $\psi_2 = \pi$ still occurs near $\chi_2 = 0$, as it does for $\alpha = 0$. The reconnection bifurcation no longer occurs at a single parameter value as it does for the integrable model \Eq{eq:1doffoldmodel} nevertheless, as can be seen in the left panel of  \Fig{fig:foldsection1}, a topological change in the the dynamics occurs just above $\chi_2 = -2.5$ where the unstable manifolds of the saddle periodic points undergo a reconnection in their intersections.

\subsection{Nearly Aligned Fold: $\alpha$ small}

When $\nu$ is nonzero but small, the dynamics of the the two degrees of freedom are coupled but $J_2$ evolves slowly compared to $J_1$. In such a situation it is appropriate to apply an averaging method. This averaging is valid when the dynamics of the first degree of freedom are far from separatrices which would cause this motion to slow.

A particularly simple situation occurs when $\alpha$ is small---indeed, as we argued in \Sec{sec:diagonalization} this is the general case. If formally $\alpha \ll \nu$, then to lowest order \Eq{eq:flow} decouples into a pendulum and the nontwist one-degree-of-freedom system \Eq{eq:1doffoldmodel}. The phase portraits then will look similar to those of \Fig{fig:foldphasespace}, just as we saw in \Fig{fig:foldsection1}

We can formally treat this case by treating $\alpha$ as a small parameter of the same order as $\nu$. Up to terms of first order in $\alpha$, \Eq{eq:flow} reduces to
\begin{align} \label{eq:alphasmallflow}
	\dot \psi_1     & =    J_1 \;, \nonumber\\
	\dot \psi_2     & =     \alpha  J_1 + \nu  ( \chi_2 +  \frac32 J_2^2) \;,\nonumber \\
	\dot J_1    & =  - a\sin\psi_1 -  \alpha b \sin\psi_2 \;, \nonumber \\
	\dot J_2    & = - \nu b \sin\psi_2 \;.
\end{align}

To lowest order we may neglect the small $\psi_2$ term in the equation for $\dot J_1$, and the $(\psi_1,J_1)$ dynamics become a simple pendulum on the fast, $O(1)$, time scale. Thus this system has periodic orbits and has true angle-action coordinates. Denoting this true angle variable by $\vartheta$ then an average over $\vartheta$ can be performed whenever its rotation frequency is large compared to $\nu$. The standard averaging theorem \cite{Lochak88} implies that for time scales $t \sim O(\nu^{-1})$ the slow averaged dynamics are given by 
\begin{align}
	\dot \psi_2     & =     \alpha  \langle J_1\rangle_\vartheta + \nu  ( \chi_2 +  \frac32 J_2^2) \;,\nonumber \\
	\dot J_2    & = - \nu b \sin\psi_2 \;.
\end{align}
where the average $ \langle \rangle_\vartheta$ is performed over  $\vartheta$. This is valid providing the first degree of freedom is not near the pendulum separatrix where the time scale separation breaks down. Barring this event, the effect on the dynamics of $(\psi_2, J_2)$ is a simple shift: instead of at $\chi_2 = 0$, the resonant saddle-center bifurcation now occurs for 
\[
    \chi_2 = - \frac{\alpha}{\nu} \langle J_1 \rangle_\vartheta
\]
The critical value $\chi^*$ for reconnection will shift down from \Eq{eq:chistar} by the same amount. However, it is important to realize that the value of $J_1$ on the Poincar\'e section is determined by $H(0,\psi_2,J_1,J_2) = E$ and so will vary with position, thus the bifurcation values corresponding to different orbits on the section will shift by different amounts. The simple prediction of a shift in the bifurcation values is verified by our numerical computations. In the example of \Fig{fig:foldsection1}, the elliptic island around $(0,J_2^{(+)})$ corresponds to $\psi_1$ in a rotating regime with $J_1$ oscillating slightly about $2.4$. Thus the predicted bifurcation point is $\chi_2 \approx -1.5$, close to the observed value of $-1.47$. By contrast the orbit around the elliptic island on the section at $(\pi,J_2^{(-)})$ has $J_1 = 1.4$ which is in the oscillating regime for the first degree of freedom so that $\langle J_1 \rangle_\vartheta = 0$. Thus the predicted bifurcation value is zero, just as observed.

The averaging theory fails when the pendulum motion in the first degree of freedom is near its separatrix. In this case the coupling between the degrees of freedom drives $J_1$ to jump across the separatrix on the slow time scale. This causes slow chaos in the second degree of freedom and will be discussed more below.





\subsection{Quasiperiodic Pendulum:  $\nu = 0$ }\label{sec:quasiperiodic}
When $\alpha = O(1)$ the two degrees of freedom of the flow \Eq{eq:flow} are coupled in an essential way; however, since $\nu \ll 1$ there is still a fast/slow spitting of the dynamics. In the extreme case that $\nu = 0$, $J_2$ is constant and is a second invariant for the flow. Be that as it may, this is a singular limit:  as we have seen earlier in this section, the dynamics with $\nu$ nonzero can result in $O(1)$ changes in $J_2$ on the time scale $\nu^{-1}$ even when $\nu$ is arbitrarily small. 

When $\nu = 0$,  \Eq{eq:flow} reduces to a Poisson system, $\dot{z} = \cJ_P \nabla H_P(z)$, with the variables $z = (\psi_1, \psi_2, J_1)$, the Hamiltonian
\[
    H_P(\psi_1, \psi_2, J_1 ) = \frac12 J_1^2 -a \cos \psi_1 -b \cos \psi_2 \;,
\]
and the Poisson matrix \[
     \cJ_P = \begin{pmatrix} 0 & 0 & \cos\alpha \\0 & 0 & \sin\alpha \\-\cos\alpha & -\sin\alpha & 0\end{pmatrix} \;,
\]
which is the upper left $3\times 3$ block of the matrix $\cJ$ in \Eq{eq:nonsymplecticForm}. This Poisson matrix has the Casimir invariant $\psi_v$ given by
\[
   \begin{pmatrix} \psi_u \\ \psi_v \end{pmatrix} = 
   \begin{pmatrix}
   \cos\alpha & \sin\alpha \\
   -\sin\alpha & \cos\alpha
   \end{pmatrix}
   \begin{pmatrix} \psi_1 \\ \psi_2 \end{pmatrix} = 
\] 
consequently, the motion is along lines of slope $\tan\alpha$ in the angle space. When the slope is irrational,  these lines are dense on the torus so the Casimir is only a local invariant.  Along a line of constant $\psi_v$ the system has infinitely many center and saddle equilibria at the points where $J_1 = 0$ and
\[
   \sin \psi_1 = -\frac{b\tan \alpha}{a}  \; \sin \psi_2\;.
\]
These occur for a set of values of the energy  $E_P = H_P$ that is dense in one or more intervals in the range $|E_P| \le |a|+|b|$. For example if $a=b>0$, then there are centers for energy values in a dense subset of $[-2a,0]$ and saddles in a dense subset of $[0,2a]$. When $E_P$ is above this range, there are no saddles, and the motion is ``untrapped."

If we define a coordinate $\psi_u$ along the line of the Casimir by $\psi_u =  \cos\alpha\; \psi_1 + \sin\alpha \;\psi_2$, this system can be obtained from the one degree-of-freedom Hamiltonian
\begin{equation}\label{eq:qpHam}
    H_{QP} (\psi_u,J_1) =  \frac12 J_1^2 - a \cos( \cos\alpha\; \psi_u - \sin\alpha\; \psi_v) - b \cos(\sin\alpha\; \psi_u + \cos\alpha \; \psi_v)
\end{equation}
with canonical equations of motion in $(\psi_u, J_1)$. However, it is important to note that $H_{QP}$ depends quasiperiodically on $\psi_u$ unless $\tan \alpha$ is rational. We show typical contours of $H_{QP}$ in \Fig{fig:quasiperiodicPlot}.

\InsertFig{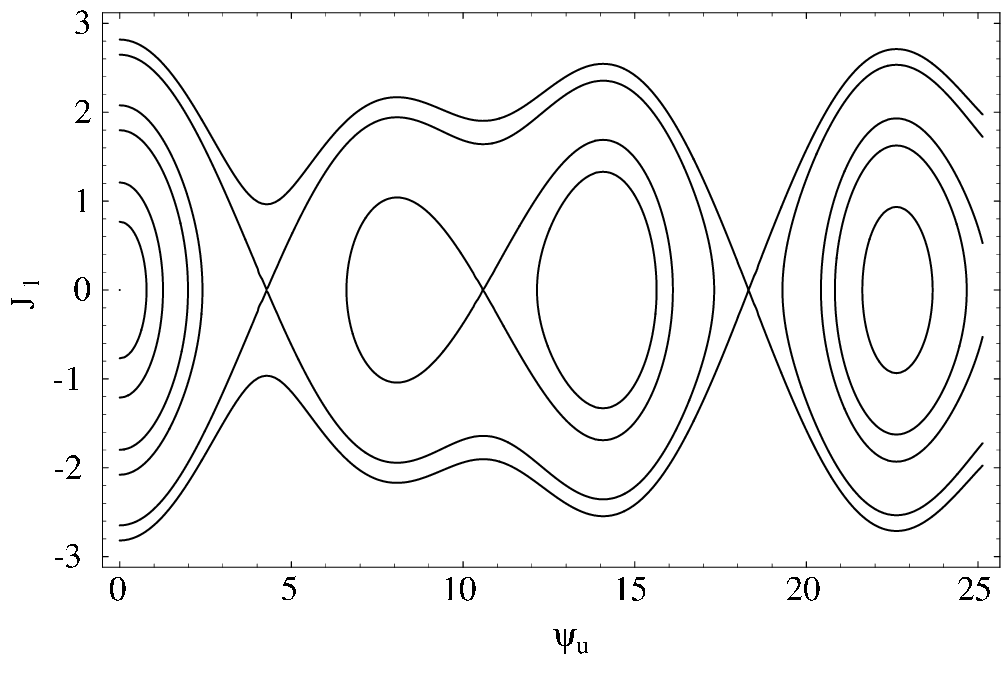}%
{Contours of the Hamiltonian \Eq{eq:qpHam} for $a=b=1$, $\psi_v = 0$ and  $\tan\alpha = \gamma$, the golden mean. Shown are the energy levels of the seven critical points of $H_{QP}$ for $\psi_u$ in the interval $[0,8\pi].$}{fig:quasiperiodicPlot}{4 in}

\subsection{Unaligned Fold: $\alpha > O(\nu)$}

The dynamics of \Eq{eq:foldHam} can be quite complex when $\alpha > O(\nu)$.  In this regime, application of averaging is problematic since the unperturbed system ($\nu = 0)$ is the quasiperiodic pendulum \Eq{eq:qpHam} instead of the simple periodic pendulum. Multi-frequency averaging is valid only when the frequencies satisfy a  Diophantine condition \cite{Lochak88}. Moreover, averaging fails for our system in the regime where the orbit passes near any of the infinitely many saddles of the quasiperiodic pendulum. Here we describe only some of the qualitative features of these dynamics.

In \Fig{fig:foldsection3} we show the effect of increasing $\alpha$ on the dynamics of \Eq{eq:flow}. In these phase portraits, $\alpha \sim 3 \nu$. The chaotic orbits in the central part of the section (near $\psi_2 = 0$) have dynamics that corresponds to multiple separatrix encounters of the quasiperiodic pendulum. There are also a number of island chains visible in this region that correspond to trapping in the potential wells of $H_{QP}$. For these islands the quasiperiodic phase $\psi_u(t)$ is bounded but ranges over intervals larger than $2\pi$. The large island near $\psi_2 = \pi$ corresponds to $(\psi_1,J_1)$ trapped near the origin, and its dynamics are relatively simple because the $\nu=0$ motion of the first degree of freedom is then periodic.  Similarly the tori for $J_2 < -1.7$ correspond to values of $J_1> 2$ where averaging over the fast quasiperiodic pendulum motion is apparently still approximately valid.

\InsertFigTwo{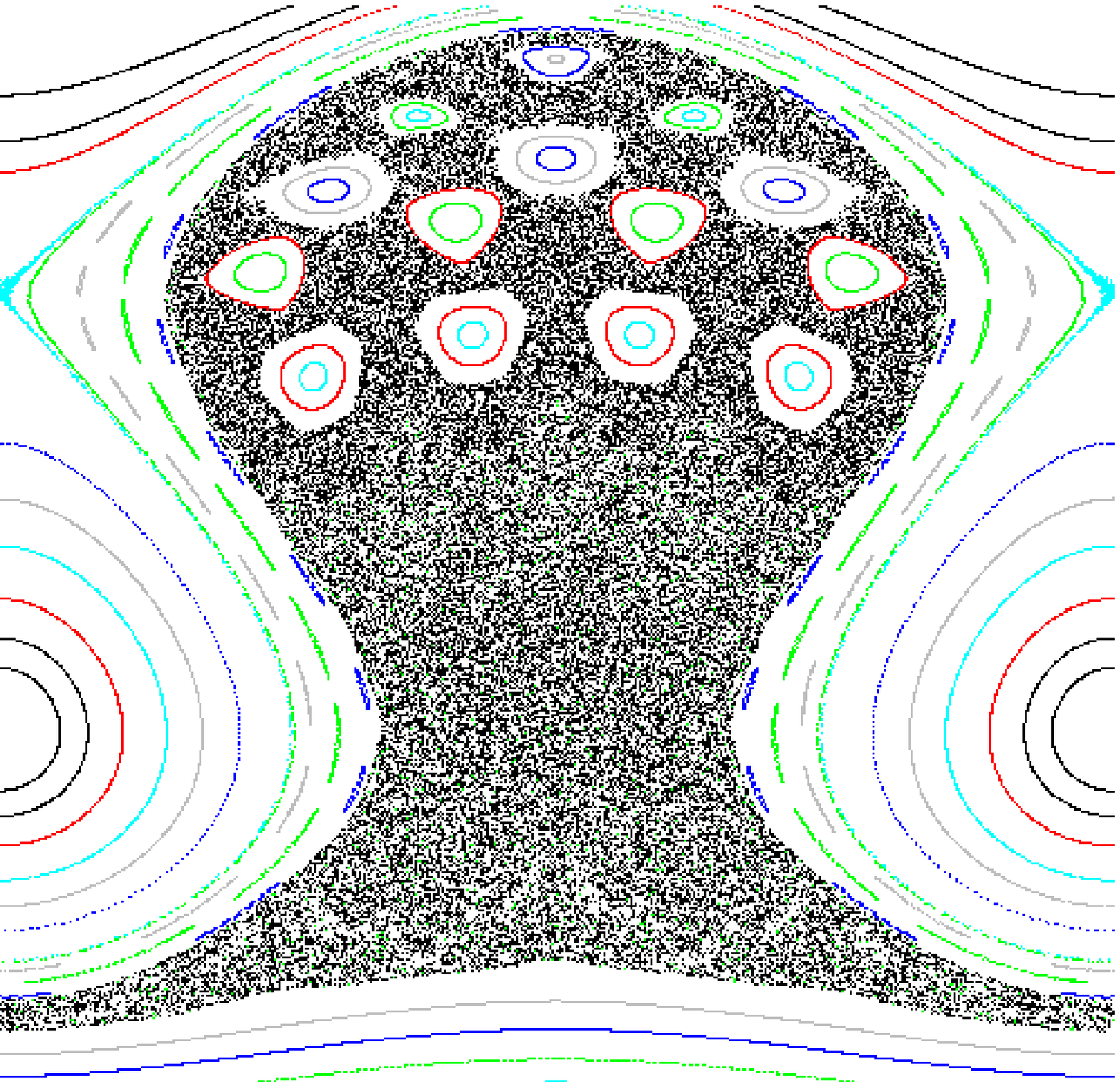}{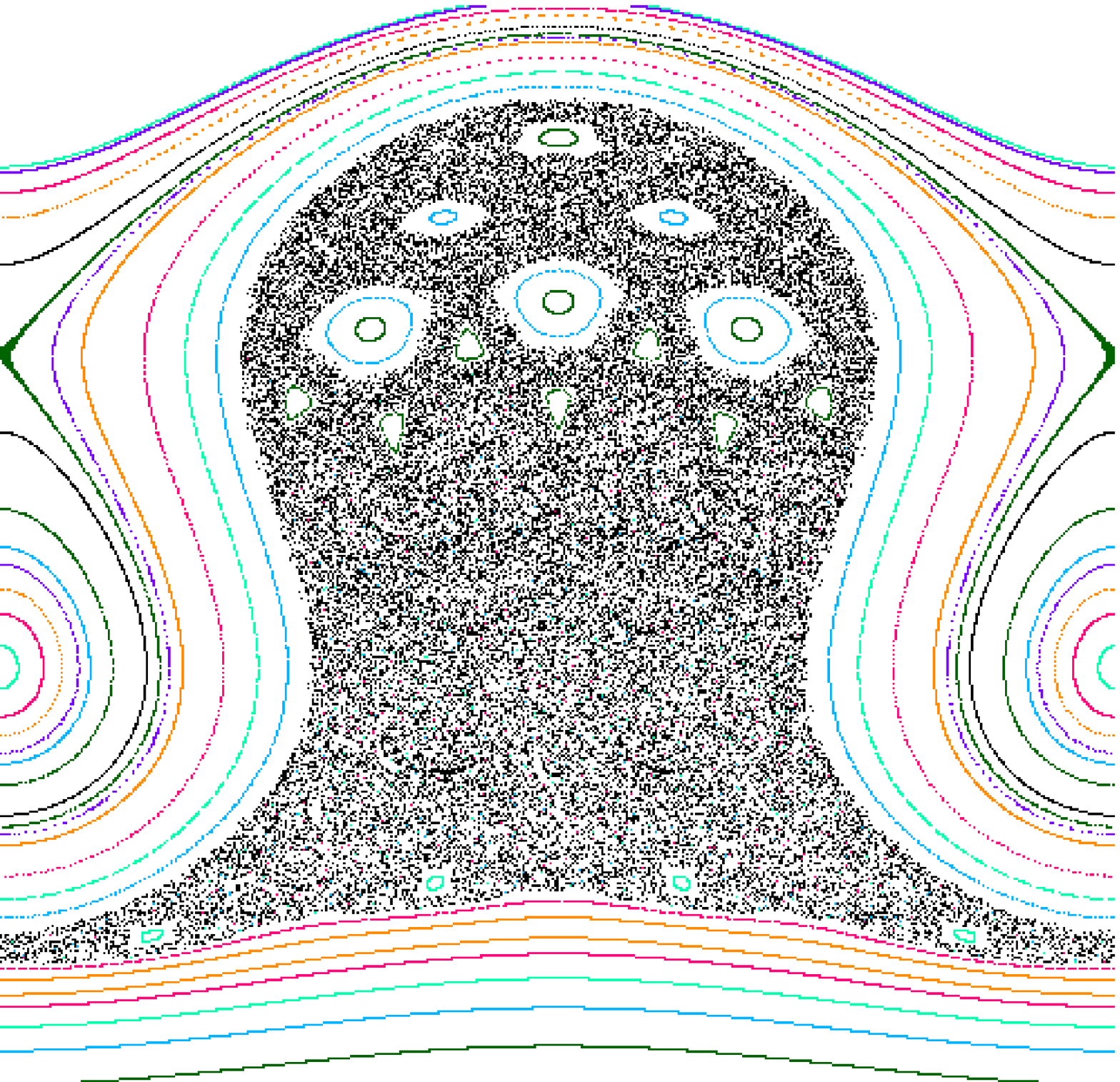}{Poincar\'e section of the flow of \Eq{eq:foldHam} for the same parameters as \Fig{fig:foldsection1} except that $\alpha = 0.05\pi$ and $E = 0.5$. For the left panel $\chi_2 = -1$ and for the right $\chi_2 = -0.5$ The vertical range corresponds to $J_2 \in \left[ -2.2, 1.8\right]$.}{fig:foldsection3}{0.4\textwidth} 

When $E < 0$ the island near $\psi_2 = \pi$ is no longer on the section---it is energetically inaccessible, see \Fig{fig:foldsection6}. In this case the invariant, meandering torus on the section boundary corresponds to the first degree of freedom undergoing small oscillations about the origin. For negative energy,  the orbits near the two fixed points $(0, J_2^{(\pm)})$ correspond to trapped motion in the first degree of freedom, and the chaotic portion of the section has moved to smaller values of $J_2$.   It is also clear from this figure that the saddle-center bifurcation of these two fixed points at $\psi_2 =0$ occurs at $\chi_2=0$, just as in the model \Eq{eq:foldHam}---a result consistent with averaging since $\langle J_1 \rangle_\vartheta = 0$ for trapped orbits.

As $E$ decreases, the energy surface intersects the section $\psi_2 = 0$ in two disconnected pieces. One corresponds to the elliptic island around $(0,J_2^{(+)})$, and the other to the nonresonant tori for $J_2 < J_2^{(-)}$. Since \Eq{eq:foldHam} is cubic in $J_2$, the energy can be arbitrarily small, but then the fold is not accessible.

\InsertFigTwo{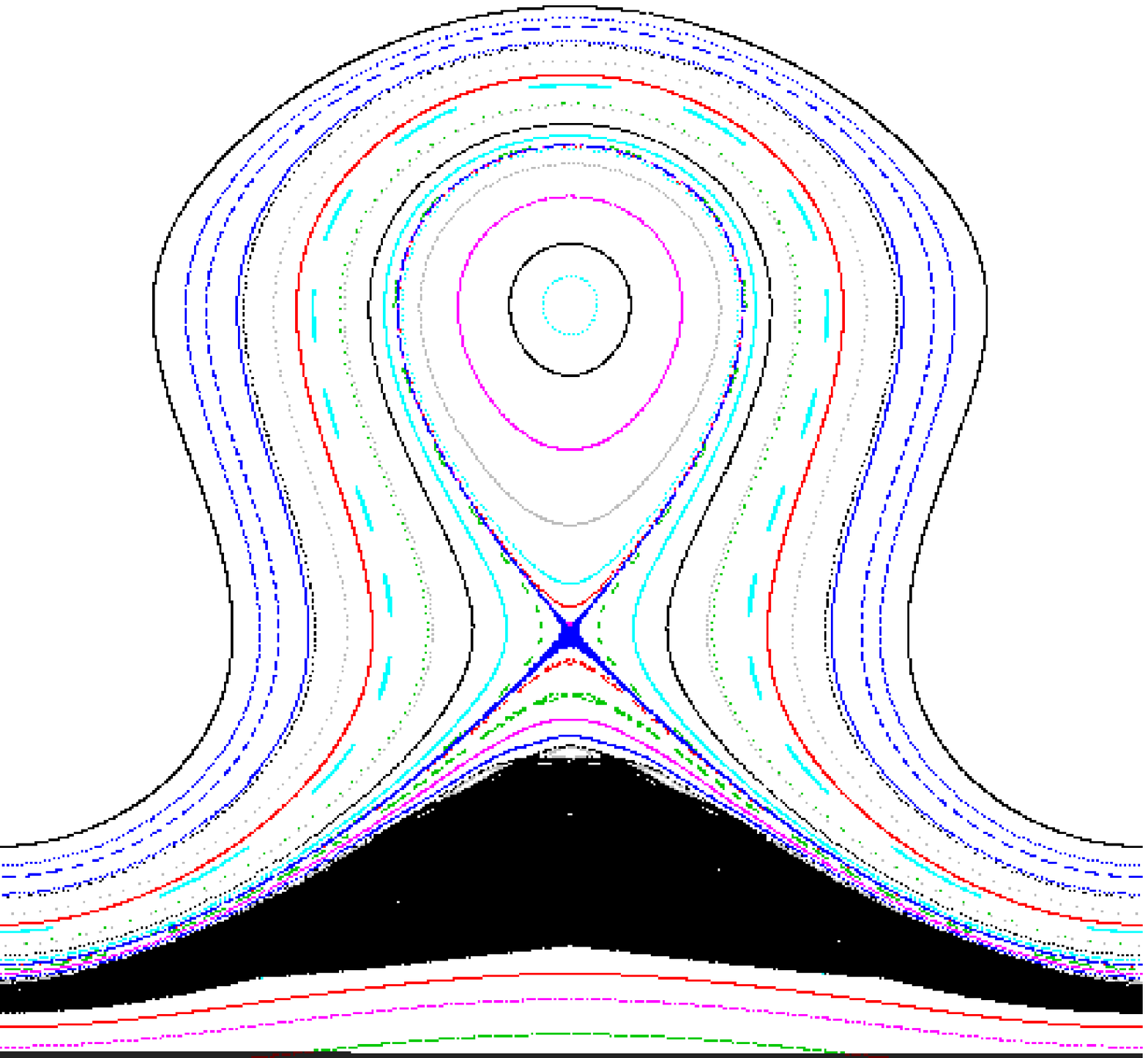}{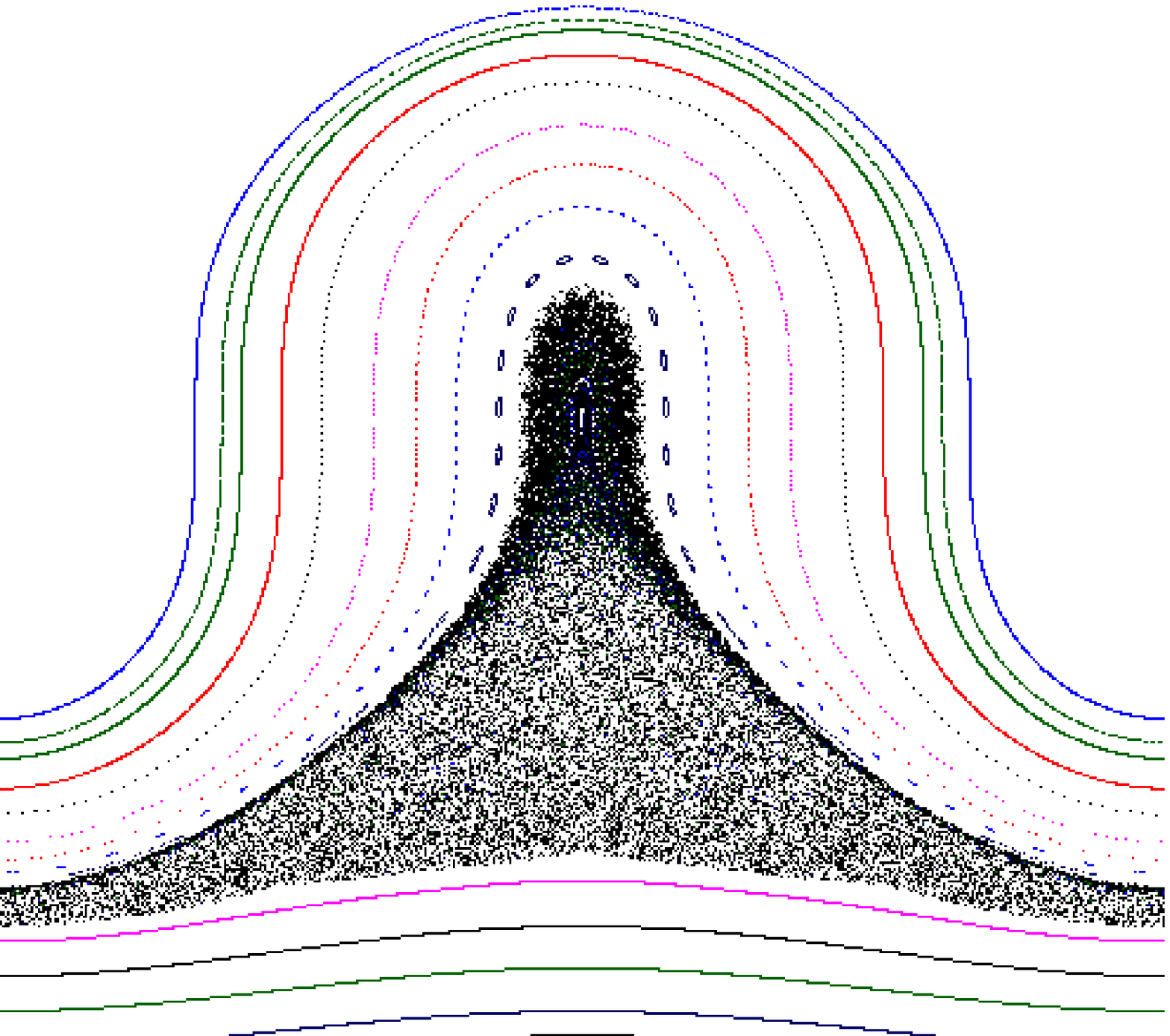}{Poincar\'e section of the flow of \Eq{eq:foldHam} for the same parameters as \Fig{fig:foldsection1} except that $\alpha = 0.05\pi$, $E = -0.5$. For the left panel $\chi_2 = -1$ and for the right $\chi_2 = 0$ The vertical range corresponds to $J_2 \in \left[ -2.2, 1.8\right]$.}{fig:foldsection6}{0.4\textwidth} 

When $\tan \alpha$ is rational, the phase portrait exhibits resonances that correspond to this coupling. An example with $\tan \alpha = \frac13$ is shown in \Fig{fig:foldsection8}. Here a period-three island chain replaces the resonance near $J_2^{(+)}$. The large degree of chaos in this section reflects orbits that repeatedly cross the separatrix of the $(\psi_1, J_1)$ pendulum. As can be seen in the right panel of this figure, as $\chi_2$ approaches $0$, there are bifurcations creating a period-two island chain as well as reconnection bifurcations of the period-three islands, which finally are destroyed in a saddle-center bifurcation at $\chi_2 = 0$. 

\InsertFigTwo{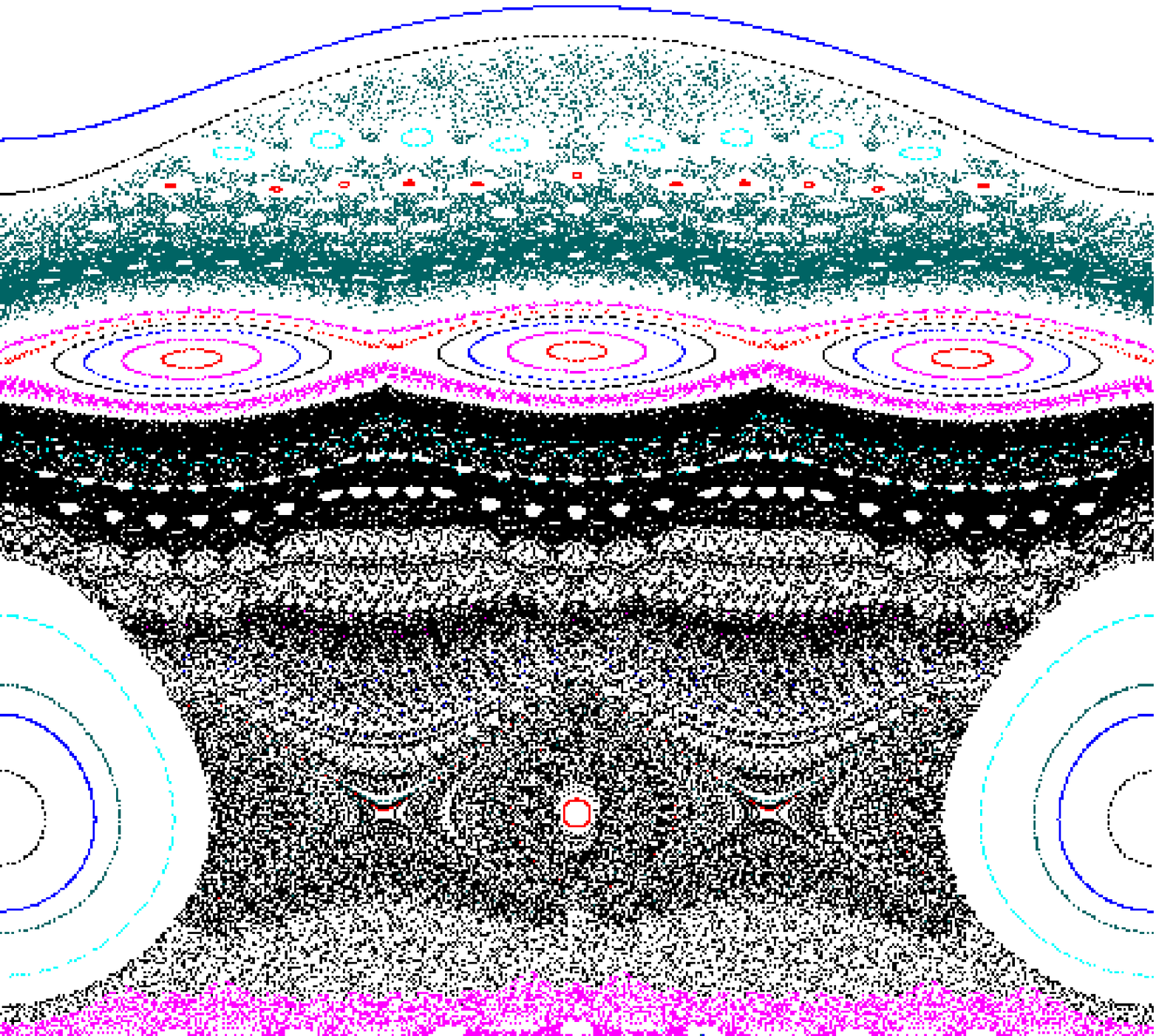}{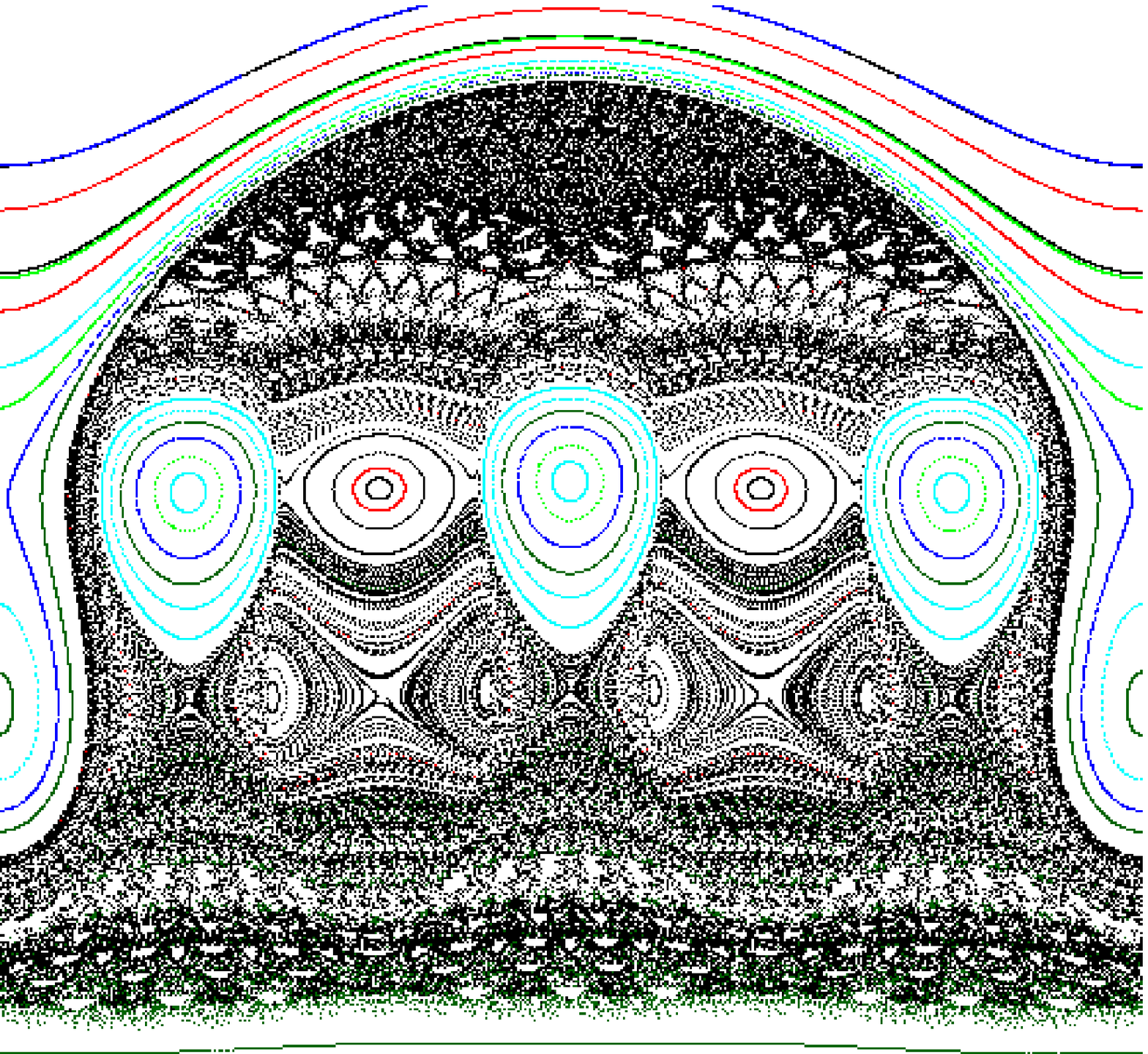}{Poincar\'e section of the flow of \Eq{eq:foldHam} for the same parameters as \Fig{fig:foldsection1} except that $\alpha = \arctan(1/3)$, $E = 1$ and $\nu = 0.01$. For the left panel $\chi_2=-1$ and the vertical range corresponds to $J_2 \in \left[ -1.6, 2.0\right]$, while for the right panel $\chi_2 = -0.15$ and $J_2 \in \left[-1.4,1.8\right]$}{fig:foldsection8}{0.4\textwidth} 

\goodbreak
\section{Cusp Model}\label{sec:cusp}

When the kinetic energy has a cusp singularity, $T = T_{cusp}$, \Eq{eq:scaledCusp} and we choose the simplest potential $\bar U$ as before, the Hamiltonian \Eq{eq:Hamiltonian} becomes
\begin{equation}\label{eq:cuspHam}
      H_{cusp}(\psi, J) =  \chi_1 J_1 + \chi_2 J_2 +\frac12 ( J_1 + \rho_0 {J}_2^2)^2 + \mu_0 {J}_2^4 
                - a \cos \psi_1 - b \cos \psi_2 \,.
\end{equation}
In this case, both components of $\chi$ are important.

\subsection{Aligned Cusp: $\alpha = 0$}

When $\alpha = 0$ the differential equations \Eq{eq:flow} become
\begin{equation} \label{eq:CUSPalpha0flow}
\begin{aligned} 
	\dot \psi_1     & =    \chi_1 + J_1 + \rho_0 J_2^2 \,,  \\
	\dot \psi_2     & =     \nu  ( \chi_2 +  2\rho_0J_2( J_1 + \rho_0 J_2^2) + 4 \mu_0 J_2^3) \,, \\
	\dot J_1    & =  - a\sin\psi_1  \,, \\
	\dot J_2    & = - \nu b \sin\psi_2 \,.
\end{aligned}
\end{equation}
The essential difference to the fold case is that for the cusp 
the equations do not separate even when $\alpha = 0$; 
both derivatives of $T_{cusp}$ depend on each of the $J_i$.
Thus, even for the case $\alpha = 0$ and small $\nu$, we
must use averaging in order to reduce to one degree of freedom.

On the fast timescale $J_2$ is constant and the equations for $(\psi_1, J_1)$ 
are that of the simple pendulum with a shifted $J_1$. 
As for the case of the fold we can introduce the angle $\vartheta$ conjugate to the true pendulum action and average over it to obtain a slow system in $(\psi_2, J_2)$ 
where $J_1$ is simply replaced by its average $\langle J_1 \rangle_\vartheta$.
A necessary condition for $J_1$ to be rotating is that the energy be sufficiently large.

After scaling time to eliminate $\nu$, the slow, averaged, one-degree-of-freedom Hamiltonian is
\begin{equation} \label{eq:H1dofCusp}
  H_{\epsilon,\beta,\chi_2}(\psi_2, J_2) = \chi_2 J_2 + \frac12  \beta J_2^2 + \epsilon J_2^4 - b \cos\psi_2 
\end{equation}
where $\beta = 2\rho_0 \langle J_1\rangle_\vartheta$ and $\epsilon = \mu_0 +  \rho_0^2/2$.
Recall from \Eq{eq:cuspgen} that the two symplectically nonequivalent forms of the cusp are represented by the sign of $\epsilon$.
Notice that negative $\epsilon$ is only possible for $\mu_0 <  0$.
By a scaling of $J_2$, $\epsilon$ can be reduced to $\pm 1$. 
The system has the discrete symmetry
\[
     H_{\epsilon,\beta,\chi_2}(\psi_2, J_2) = -H_{-\epsilon,-\beta, \chi_2}(\psi_2+\pi, -J_2)
\]
so that for the bifurcation analysis it is enough to consider, e.g.,  $\epsilon = +1$. The bifurcation diagram as a function of the two unfolding parameters $\chi_2$ and $\beta$ for this case is shown in \Fig{fig:bifdiag}. For $\epsilon = -1$ the picture is reflected 
through the vertical axis so that the critical values for $\beta$ all become non-negative.
There are four bifurcation curves shown in \Fig{fig:bifdiag}. 
The curve $ \beta^3 + 27 \chi_2^2 = 0$ with a cusp at the origin corresponds to 
a saddle-center bifurcation. The three remaining curves mark reconnection bifurcations.%
\footnote
{
	The reconnection curves are given by
	$
	-4194304 + 8192 \beta^4 - 4\beta^8 - 1769472\beta\chi_2^2 
	+ 1728\beta^5\chi_2^2 - \beta^9\chi_2^2 - 186624\beta^2\chi_2^4 
	      - 81\beta^6\chi_2^4 - 2187\beta^3\chi_2^6 - 19683\chi_2^8 = 0
	$.
}
The lower line of phase portraits all correspond to $\chi_2 = 0$;
they are symmetric under $J_2 \to -J_2$.
The shown part of the bifurcation diagram with $\chi_2 \ge 0$ has four disjoint regions
separated by curves of critical values.
The lower half of the bifurcation diagram with $\chi_2 < 0$ is related to 
the upper half shown in \Fig{fig:bifdiag} by the discrete symmetry 
\[
     H_{\beta,\chi_2}(\psi_2, J_2) = H_{\beta, -\chi_2}(\psi_2, -J_2) \,.
\]

\InsertFig{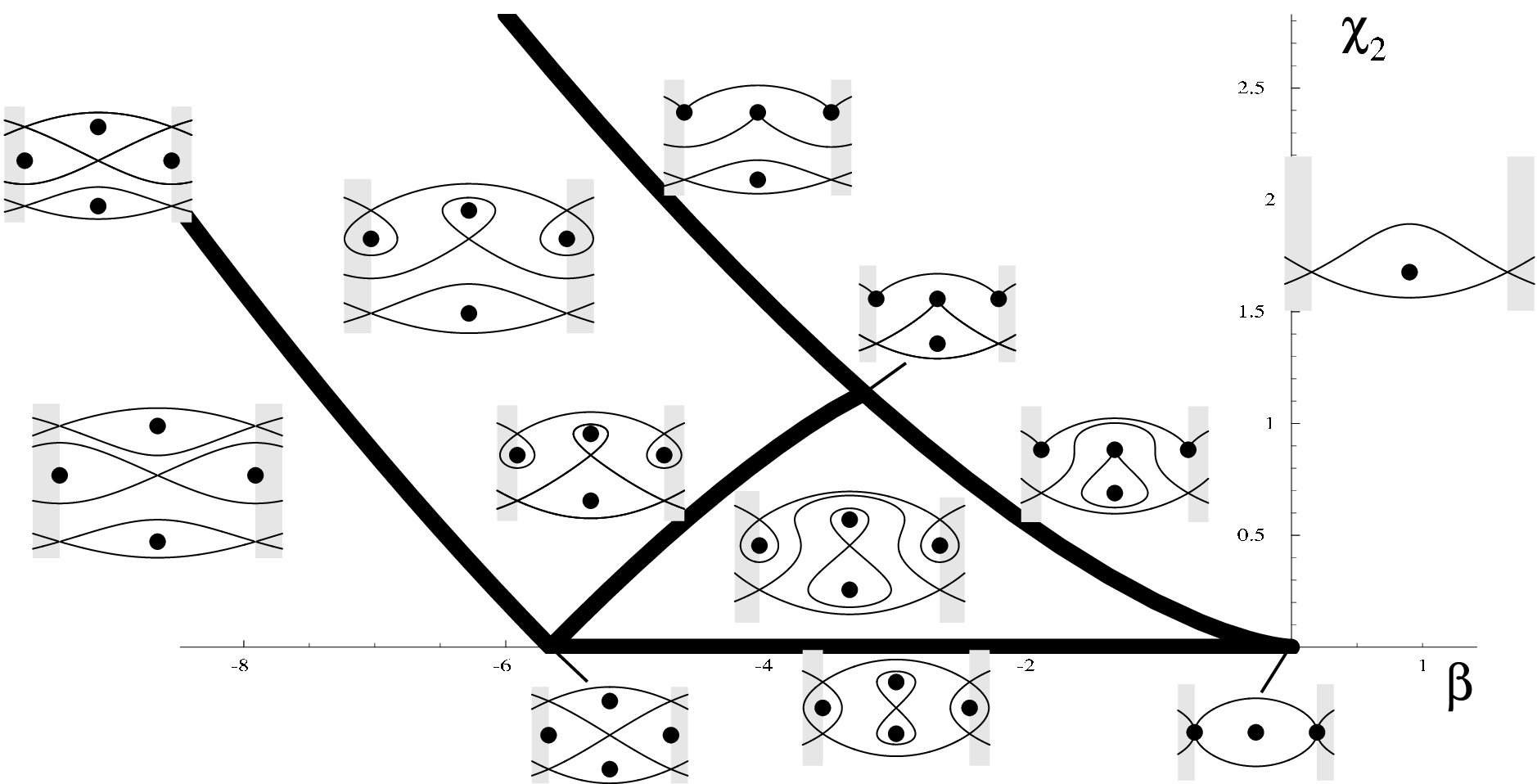}{Bifurcation diagram of the averaged one-degree-of-freedom 
Hamiltonian \Eq{eq:H1dofCusp} with $\epsilon = +1$. 
The dots in the phase portraits mark isolated stable or bifurcating equilibrium points. 
There are four generic phase portraits inside the respective regions.
The five codimension one phase portraits are touching the fat critical lines.
The three codimension two points are at 
$(\beta, \chi_2) = (0,0), (-4\sqrt{2},0), (-4\sqrt{2/3}, (8/3)^{1/4} 8/9)$.}%
{fig:bifdiag}{0.8\textwidth} 

Moving counter clockwise on a large circle around the origin starting in the 4th quadrant 
near $\beta = 0$ reproduces the twistless bifurcation associated to a fold: first the creation of two island chains in a saddle-center bifurcation and next the reconnection.
The characteristic feature of the cusp appears close to the origin in parameter space
where the lines corresponding to folds are close to each other. 
The three reconnections lines differ in which of the three different lines
of equilibria is reconnecting. A qualitatively new phase portrait appears in the 
triangular region attached to the origin, which has a localized invariant twistless 
curve. The main feature of this new region is that there are twistless tori which are not
rotational invariant circles between the figure-eight separatrix and the rotational separatrix.
Unlike the meandering curves (which are also present in the same phase portrait) 
these are, however, localized in $\psi_2$.

Each of these regimes is easily visible in numerical phase portraits if one chooses parameter values so that the motion of the first degree of freedom gives rise, through $\avJ$, to an appropriate value for $\beta$. In  Figs. \ref{fig:cuspsection1}-\ref{fig:cuspsection2} we choose $\mu_0 > 0$ and $\rho_0 < 0$ so that $\epsilon = +1$, and fix all of the other parameters of \Eq{eq:cuspHam}. The dynamics of the first degree of freedom is then determined by the value of the energy: when $E$ is large, points on the Poincar\'e section have large positive $\avJ$, which makes $\beta$ large and negative. As $E$ decreases, so does $\avJ$, moving the phase portrait horizontally in the bifurcation diagram \Fig{fig:bifdiag}. Each of the four regimes of \Fig{fig:bifdiag} are shown in Figs. \ref{fig:cuspsection1}-\ref{fig:cuspsection2}. The degree of chaos in these panels also increases, to the final case where the dynamics of $J_1$ undergoes separatrix crossings over a significant portion of the phase space.

\InsertFigTwo{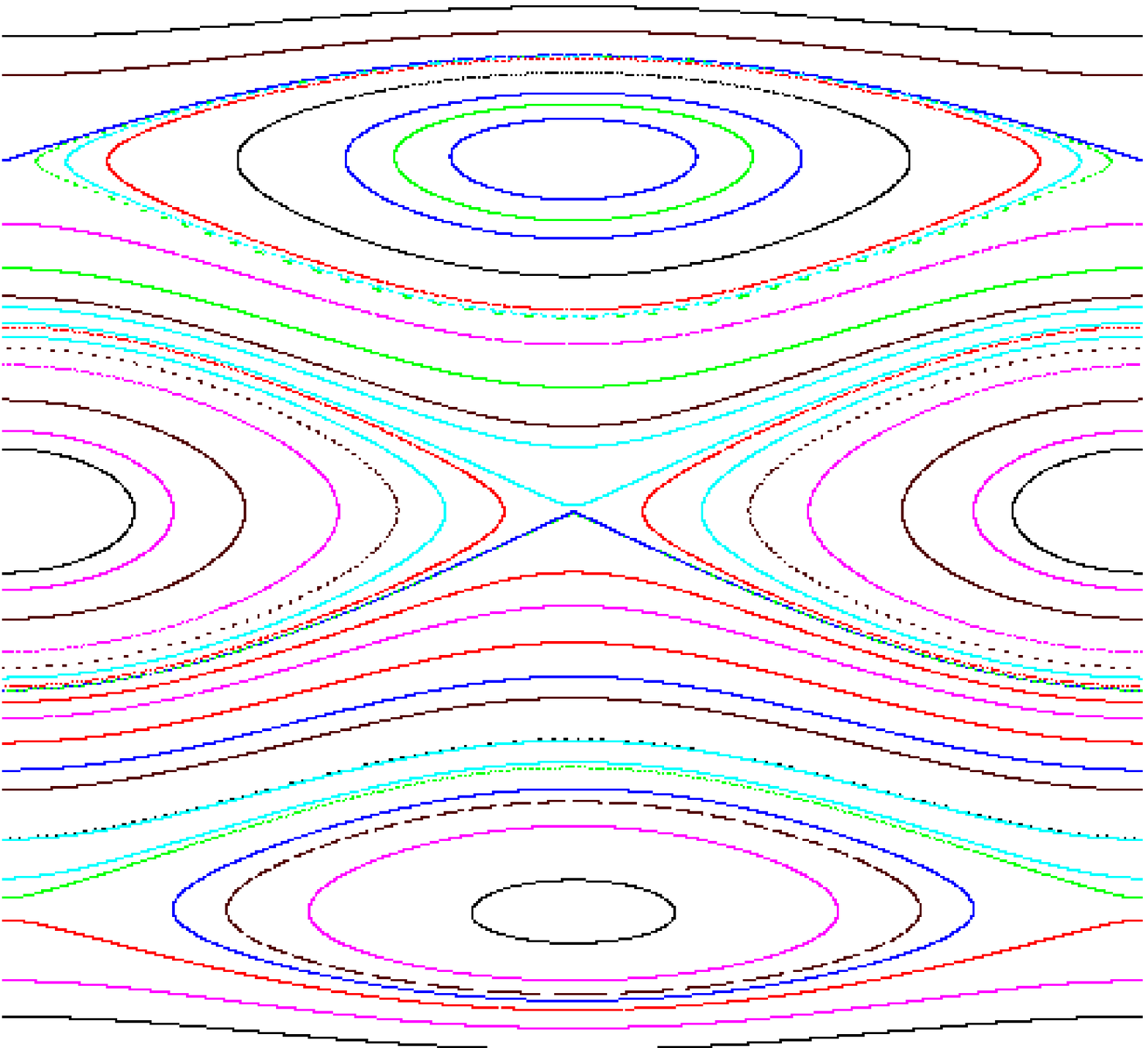}{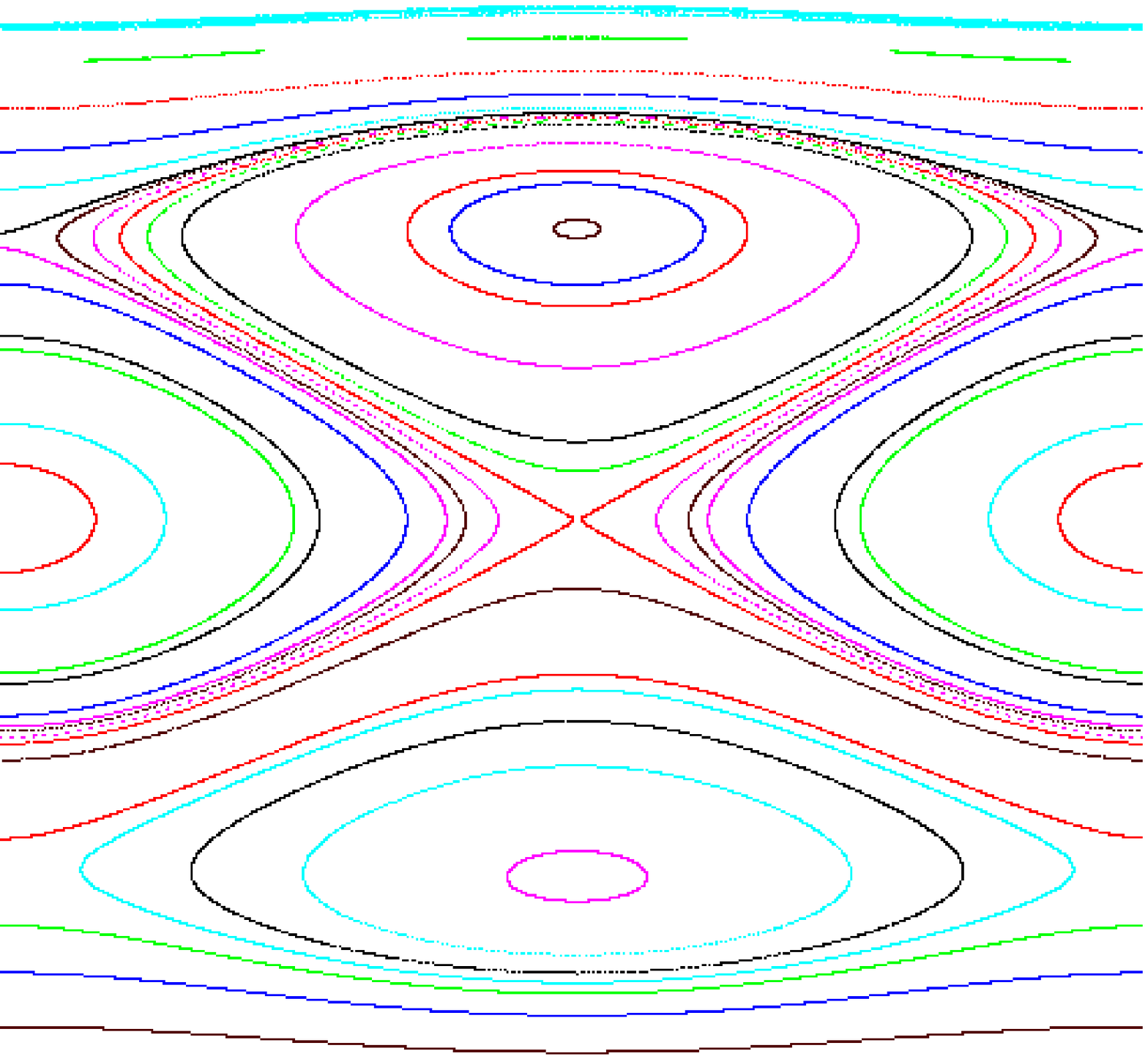}{Poincar\'e sections of the cusp Hamiltonian \Eq{eq:cuspHam} with parameters $\rho_0 = -0.5$, $\mu_0 = 0.875$, $a = b = 1$, $\nu = 0.05$, $\alpha =0$, $\chi_1 = 0$ and $\chi_2 = 0.5$. For the left panel $E = 35.75$, and for the right $E = 18.11$. The vertical range is $J_2 \in \left[ -0.2, 1.0\right]$. The energy determines the value of $\avJ$.}{fig:cuspsection1}{0.4\textwidth} 

\InsertFigTwo{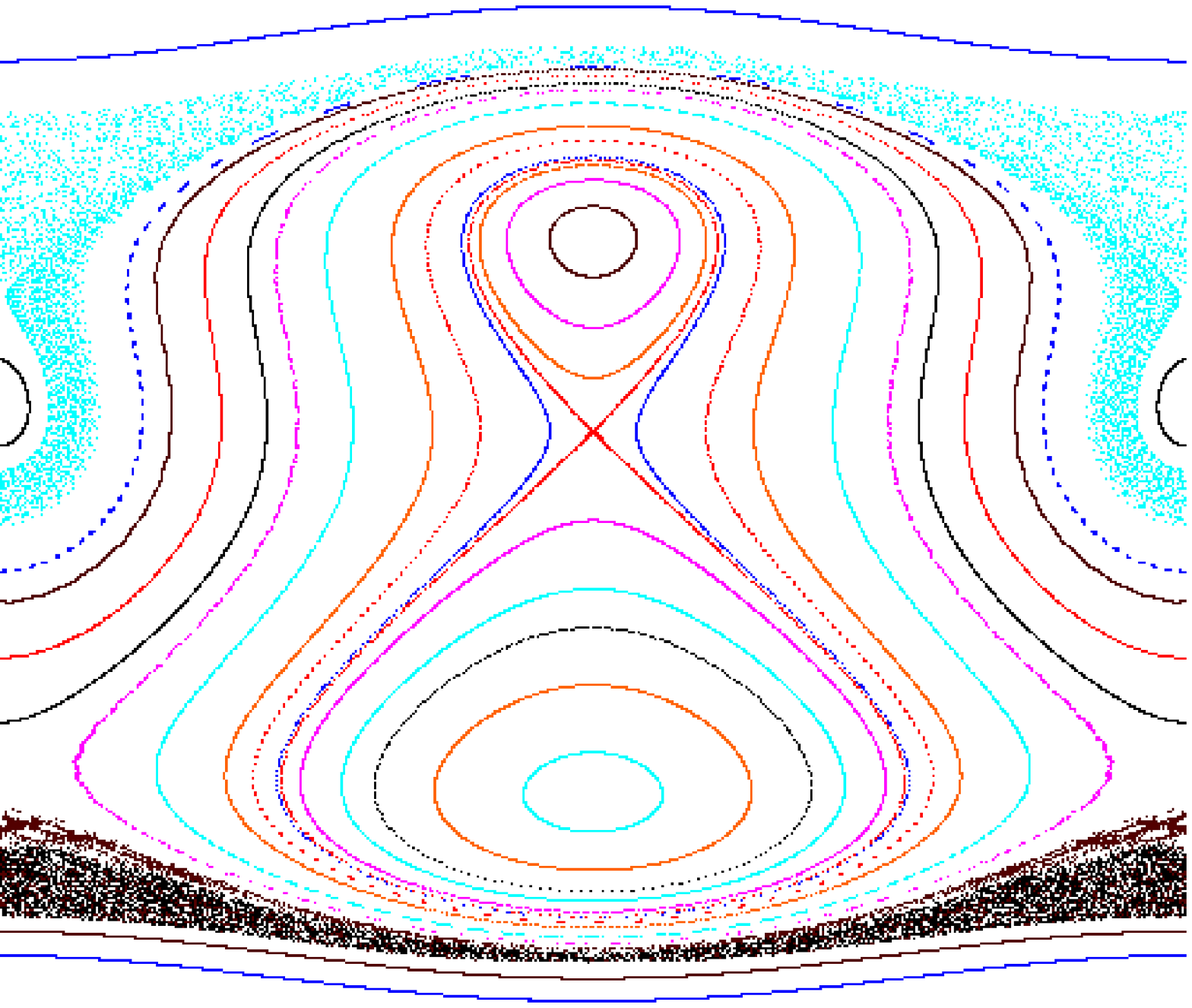}{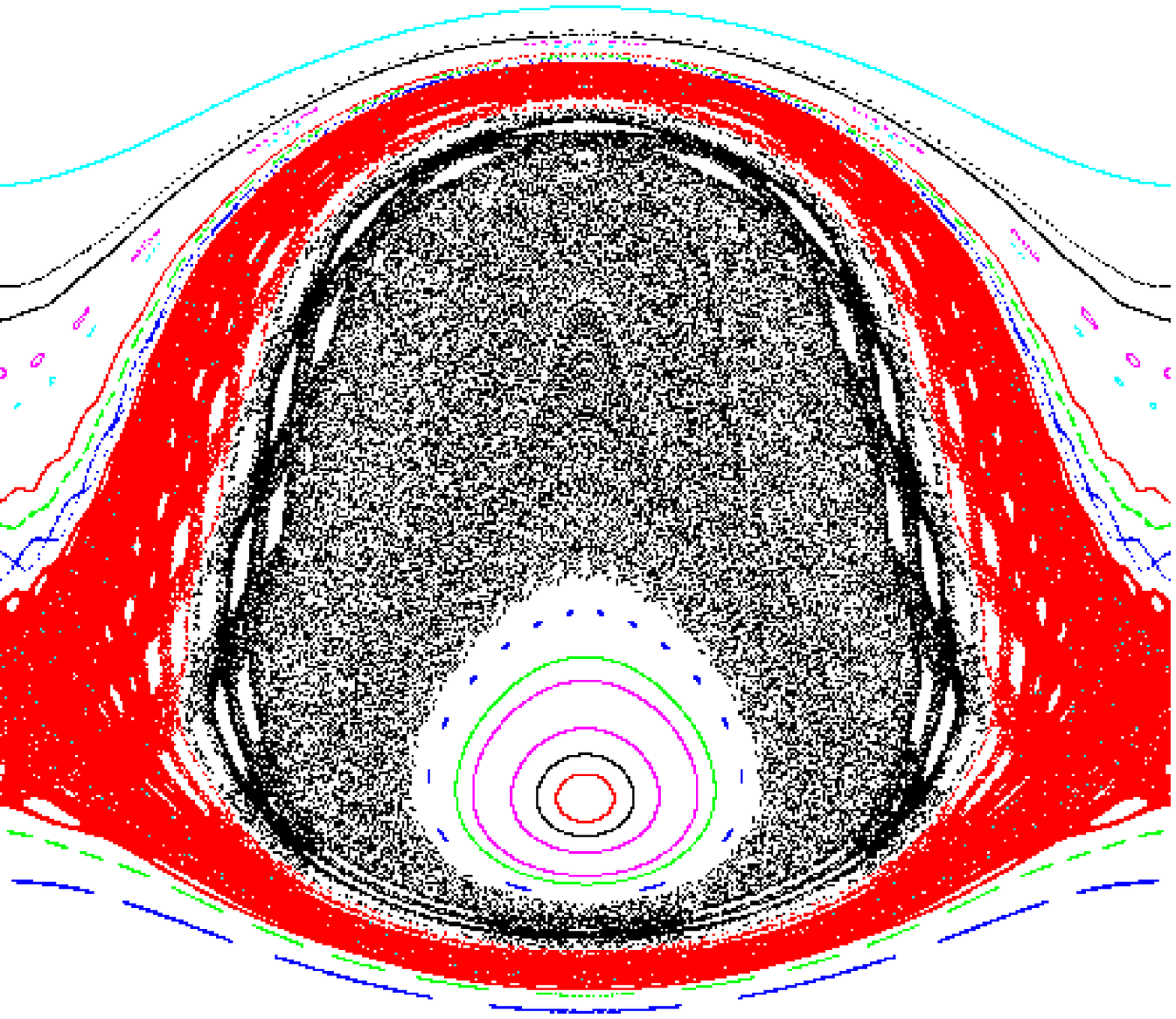}{Poincar\'e sections of the cusp Hamiltonian \Eq{eq:cuspHam} for the same parameters as \Fig{fig:cuspsection1}. For the left panel $E = 3.75$, and for the right $E = 0.75$. The vertical range is $J_2 \in \left[ -1.7, 1.6\right]$.}{fig:cuspsection2}{0.4\textwidth} 
\subsection{Nearly Aligned Cusp: Small $\alpha$}

Treating $\alpha$ to be of the same order as the small parameter $\nu$, the first order approximation is
\begin{equation} \label{eq:CUSPalphasmallflow}
\begin{aligned} 
	\dot \psi_1     & =    \chi_1 + J_1 + \rho_0 J_2^2 \,,  \\
	\dot \psi_2     & =     \nu  ( \chi_2 +  2\rho_0J_2( J_1 + \rho_0 J_2^2) + 4 \mu_0 J_2^3) +
		\alpha(\chi_1 + J_1 + \rho_0 J_2^2) \,, \\
	\dot J_1    & =  - a\sin\psi_1  - \alpha b \sin \psi_2  \,, \\
	\dot J_2    & = - \nu b \sin\psi_2 \,.
\end{aligned}
\end{equation}
The Hamiltonian for these equations again is \Eq{eq:cuspHam}, but the 
equations of motion are different from the $\alpha = 0$ case because of the 
non-diagonal non-standard symplectic form.
Averaging works as before after neglecting the small $\psi_2$ term in the equation 
for $\dot J_1$. After averaging the equations of motion, we see they are generated by 
the one-degree-of-freedom Hamiltonian in $(\psi_2, J_2)$:
\[
    H = \left(\chi_2 + \frac{\alpha}{\nu}( \chi_1 + \avJ)\right) J_2 + \rho_0 \avJ J_2^2 + 
    \frac{\alpha}{3\nu} \rho_0 J_2^3 +(\mu_0 + \rho_0^2/2) J_2^4 - b \sin\psi_2 \,.
\]
Finally this Hamiltonian can be recast in the form \Eq{eq:H1dofCusp}
by completing the quartic, scaling, and collecting terms so that new effective
parameters $\beta$ and $\chi_2$ are found.
Denote the change of scale of $J_2$ necessary in order to make the quartic 
coefficient equal to $\epsilon  = \pm 1$ by $s$, $s^4 = \epsilon/(\mu_0 + \rho_0^2/2)$.
Then the shift in $J_2$ is $\lambda = -s^3 \epsilon \alpha \rho_0/(12\nu)$. 
The shift in $\beta = 2 \avJ \rho_0 s^2$ is given by 
$ - 6\epsilon \lambda^2$.
Finally $\chi_2$ is shifted by 
\[
   s \left( \chi_1 + \avJ \frac{\alpha}{\nu}\right)
 + 2 \lambda s^2 \rho_0 \avJ
   - 8 \epsilon \lambda^3
\,.
\]
Since when $\alpha/\nu = {\cal O}(1)$, the shift $\lambda$ is small, the dominant term in these shifts is typically the one involving $\avJ$. 
Thus in particular making $\alpha$ non-zero breaks the discrete 
symmetry present when $\chi_2 = 0$.

\subsection{Quasiperiodic case: $\nu = 0$}

Setting $\nu = 0$ causes $J_2$ to be a constant, and thus again we obtain 
a quasiperiodic system with one degree of freedom. 
The essential difference to the fold case \Eq{eq:qpHam} is that here there is a linear
term in $J_1$ in the Hamiltonian. 
Thus up to a shift of $\chi_1 + \rho J_2^2$ in $J_1$ again the quasiperiodic 
pendulum is found.

Because the critical values are dense in an interval the quasiperiodic pendulum also
has an abundance of twistless curves. In any region of non-critical points 
in phase space bounded by the separatrices of two distinct hyperbolic points there is
a twistless curve.



\section{Conclusion}
 
 We have derived two normal forms for a nearly integrable, four-dimensional symplectic map with a singularity in its frequency ratio map, $J \mapsto \Omega(J)$.  A singularity corresponds to the vanishing of the twist $\tau = \det(D\Omega)$. For a four-dimensional mapping there are two stable singularities: the fold and cusp. In each case we focus upon a point  $\Omega = \omega^*$ where the caustic crosses a resonance curve $\langle  m, \Omega(J) \rangle \in \Z$ for $m \in \Z^2$.  We defined coordinates $(J_1, J_2)$ so that the $J_1$-axis is tangent and the $J_2$-axis is normal to the singular set. Consequently, $J_2$ represents the degenerate action and a singularity in the frequency map then corresponds to the vanishing of quadratic terms in the kinetic energy for $J_2$. Using $\epsilon$ to denote the size of the perturbation, the dominant balance along the singular set gives $J_1^2 \sim O(\epsilon)$. If the singularity is a fold, then dominant balance transverse to the singular set gives $J_2^3 \sim O(\epsilon)$, while if it is a cusp then $J_2^4 \sim O(\epsilon)$. The differential scaling gives rise to the small parameter $\nu = \epsilon^{1/6}$ or $\epsilon^{1/4}$, respectively in our normal form.
 
Because it is nearly resonant, the normal form has an approximate invariant---an energy---that restricts the motion to three-dimensional surfaces. The corresponding dynamics is given by the flow of a two-degree-of-freedom Hamiltonian system of the form $H = \langle \chi, J \rangle + T(J) + U(\psi)$.  As can be expected, this system is generally nonintegrable. One source of nonintegrability is standard: the potential $U$ contains terms that couple the degrees of freedom. Indeed this occurs even in the case of nondegenerate frequency maps, where $T(J) = \frac12 (J_1^2 + J_2^2)$, \cite{Poschel93}.

In our model, however, there are two additional sources of complexity. First, the cusp model has an intrinsic coupling between the longitudinal ($J_1$) and transverse ($J_2$) actions because the kinetic energy $T$ in  \Eq{eq:scaledCusp} is not separable. 
The second source corresponds to the fact that the singular set is not generally aligned with the resonance module, and indeed will generically have an irrational slope. This gives rise to the misalignment angle $\alpha$ in \Eq{eq:cDefine}, and the corresponding nonsymplectic structure of our equations of motion \Eq{eq:flow}. Though it is common for issues of rationality to be important in Hamiltonian dynamics for rotation numbers, as far as we know, the rationality of a slope in action space has not been important in any other situation.

Because of the disparate scaling in the action variables, the time scale for evolution of the degenerate action $J_2$ and its conjugate angle are slower by the factor $\nu$ than those for the nondegenerate ones. If we simply set $\nu$ to zero, then $J_2$ is constant. The dynamics for this case reduced to 
a system with one degree of freedom with a quasiperiodic potential. An equivalent description 
of this quasiperiodic pendulum uses a rank one Poisson-structure on $\R \times \T^2$.

The simplest dynamics occurs for the fold when $\alpha = 0$, when the fold curve is parallel to a vector in the resonance module, and when the coupling harmonics in $U$ are negligible. For this case the multi-dimensional fold behaves as the well-known one-degree-of-freedom case derived by Simo \cite{Simo98}. However, even the fold has complex behavior when $\alpha \neq 0$. We have seen that the coupling of the nonsingular action $J_1$ to the degenerate degree of freedom $J_2$ leads to slow chaos when the first degree of freedom is near a separatrix. It also causes the shift in the reconnection bifurcation values when $J_1$ is in a rotating regime.

The cusp dynamics corresponds to the collision of two fold curves. Consequently, near a cusp there can be three sets of island chains with the same rotation number. As the cusp is approached there are several ``reconnection" scenarios. The intrinsic coupling of the two degrees of freedom in this case make the study of this system more complex. Even when $\alpha = 0$ the two degree-of-freedom 
normal form flow is non-integrable. When $\nu$ is small averaging allows for reduction to one 
degree of freedom. This simple system has been completely analyzed, and good agreement is
found with the full equations for appropriate choice of parameters. 
A new type of twistless torus which is not a graph over the unperturbed torus appears near the cusp.
Similar to the case of the fold the case of small $\alpha$ can again be reduced to the same one degree of freedom model, with shifted parameters, however.

It would be interesting to study the non-integrable dynamics of the normal form flow with two
degrees of freedom. It seems possible to still use averaging theory to partially reduce 
the dynamics, but the crossing of separatrices and the corresponding jumps in action 
would need to be taken into account. Another interesting case is the perturbation 
away from the quasiperiodic case $\nu = 0$ for arbitrary $\alpha$ using some kind of
quasiperiodic averaging theory.
  
\clearpage
\appendix
\section{Appendix: Nonsymplectic Coordinates}\label{sec:nonsymplectic}
 
A canonical generating function $S(\vphi, I')$ implicitly defines a symplectic map by the equations
\[
   \vphi ' =  \frac{\partial}{\partial I'} S \;, \quad
     I    =     \frac{\partial}{\partial   \vphi} S \;.
\]
This map preserves the symplectic form 
\[
   \varpi = d I \wedge d \vphi \;,
\]
since $\varpi - \varpi' = d(I d \vphi - I'  d \vphi' ) = d(I d \vphi + \vphi' d I' ) = d^2 S =  0$. The symplectic form has the equivalent matrix representation
\[
   \varpi =     \begin{pmatrix} 0 & -id \\ id & 0  \end{pmatrix} \;,
\]
 in $(\vphi,I)$ coordinates.
 
Under a general, linear point transformation $(\psi,J) = (B\vphi, C I)$, we can define a new generating function
\[
	\hat{S}(\psi,J') = S(B^{-1} \psi, C^{-1} J') \;.
\]
The dynamically equivalent map in the new, nonsymplectic coordinates is

\[
     \psi' = BC^t \frac{\partial}{\partial J'} \hat{S} \;, \quad
     J    =  CB^t § \frac{\partial}{\partial   \psi} \hat{S} \;.
\]
Thus the transformation is symplectic if $B=C^{-t}$.

Symplectic transformations with multiplier are a special case of the above 
where $BC^t$ is a multiple of the identity.

\bibliographystyle{alpha}
\bibliography{BibFile}
\end{document}